\documentclass[fleqn,usenatbib]{mnras}
\usepackage{txfonts}
\usepackage{multirow}
\usepackage{lineno}
\usepackage{graphicx}
\usepackage{xcolor}
\usepackage{pdflscape}
\usepackage[T1]{fontenc}    % use 8-bit T1 fonts
\usepackage{subfig}
\usepackage{ulem}

\usepackage{url}

\usepackage{setspace}

\usepackage{slashbox}
\newcommand{\gsim}{{\lower.5ex\hbox{$\; \buildrel > \over \sim \;$}}}
\newcommand{\ergs}{erg cm$^{-2}$ s$^{-1}$}
\title[Deep Learning Classification of Blazars]{Deep Learning Blazar Classification based on Multi-frequency Spectral Energy Distribution Data}

\author[Fraga et al.]
{Bernardo M. O. Fraga$^1$,\thanks{bernardo@cbpf.br}
Ulisses Barres de Almeida$^1$, 
Cl\'{e}cio R. Bom$^{1,2}$, 
Carlos H. Brandt$^3$,
\newauthor Paolo Giommi$^{4,5,6}$,
Patrick Schubert$^{1}$
and M\'{a}rcio P. de Albuquerque$^1$\\
$^1$ Centro Brasileiro de Pesquisas F\'isicas, Rua Dr. Xavier Sigaud 150, 22290-180 Rio de Janeiro, RJ, Brazil\\
$^2$ Centro Federal de Educa\c{c}\~{a}o Tecnol\'{o}gica Celso Suckow da Fonseca, Rodovia M\'{a}rcio Covas, lote J2, quadra J - Itagua\'{i} (Brazil)\\
$^3$ Jacobs University Bremen gGmbH, Campus Ring 1, 287950 Bremen, Germany\\
$^4$ Agenzia Spaciale Italiana (ASI), Via del Politecnico snc, 00133 Roma, Italy\\
$^5$ Excellence Cluster ORIGINS, Boltzmannstrasse 2, D-85748 Garching bei M\"unchen, Germany \\
$^{6}$ Center for Astro, Particle and Planetary Physics (CAP3), New York University Abu Dhabi, PO Box 129188 Abu Dhabi, United Arab Emirates;\\
}
\date{Submited XXX}

\pubyear{2020}

\begin{document}
\label{firstpage}
\pagerange{\pageref{firstpage}--\pageref{lastpage}}
\maketitle
% FRONTMATTER
%\begin{frontmatter}

% ====================================================================
% Abstract
% ====================================================================
\begin{abstract}

Blazars are among the most studied sources in high-energy astrophysics as they form the largest fraction of extragalactic gamma-ray sources and are considered prime candidates for being the counterparts of high-energy astrophysical neutrinos. Their reliable identification amid the many faint radio sources is a crucial step for multi-messenger counterpart associations. As the astronomical community prepares for the coming of a number of new facilities able to survey the non-thermal sky at unprecedented depths, from radio to gamma-rays, machine learning techniques for fast and reliable source identification are ever more relevant. The purpose of this work was to develop a deep learning architecture to identify blazar within a population of AGN based solely on non-contemporaneous spectral energy distribution information, collected from publicly available multi-frequency catalogues. This study uses an unprecedented amount of data, with SEDs for $\approx 14,000$ sources collected with the Open Universe VOU-Blazars tool. 
It uses a convolutional long-short term memory neural network purposefully built for the problem of SED classification, which we describe in detail and validate. The network was able to distinguish blazars from other types of AGNs to a satisfying degree (achieving a ROC area under curve of $0.98$), even when trained on a reduced subset of the whole sample. This initial study does not attempt to classify blazars among their different sub-classes, or quantify the likelihood of any multi-frequency or multi-messenger association, but is presented as a step towards these more practically-oriented applications.

\end{abstract}

% ====================================================================
% Keywords
% ====================================================================
\begin{keywords}
galaxies: active -- BL Lacertae objects: general -- methods: data analysis -- astronomical data bases: miscellaneous -- virtual observatory tools

\end{keywords}

%\end{frontmatter}

% ====================================================================
% Introduction
% ====================================================================

\section{Introduction}
\label{sec:introduction}

Blazars, a small radio-loud subclass of Active Galactic Nuclei (AGN), constitutes the dominant population of extragalactic sources in the gamma-ray sky. Blazars comprise 3,413 out of the 5,787 sources in the fourth Fermi Large Area Telescope catalog (4FGL)~\citep{4FGL, 4FGLDR2}, and 98\% of the extragalactic sources in the Third Catalog of Hard Fermi-LAT Sources (3FHL)~\citep{3FHL}, which collects sources detected above 10 GeV. They also account for the large majority (73 out of 84\footnote{http://tevcat2.uchicago.edu/}) of the currently known very-high energy (VHE) extragalactic sources detected by ground-based gamma-ray instruments and are expected to be the dominant type of sources detected by the future Cherenkov Telescope Array at high Galactic latitudes.
~\citep{CTAbook}. Blazars emit variable, non-thermal emission throughout the electromagnetic spectrum, and their spectral energy distribution (SED) -- a plot of the energy flux $\nu F_{\nu}$ versus frequency $\nu$ -- is formed by two broad bumps attributable, at lower energies, to synchrotron radiation and, at higher energies, to inverse-Compton emission~\citep{1998MNRAS.299..433F,AGNReview}. Their numerical prevalence in high-energy catalogues is a result of the sources' extreme emission properties, which is dominated by relativistic beamed radiation from a jet closely aligned to the observer's line of sight.

Blazars were originally distinguished from other AGN on the basis of their rapid variability or lack (sometimes, weakness) of emission lines in their optical spectrum. 
The population is divided into BL Lacs and Flat Spectrum Radio Quasars (FSRQs) depending on the strength of their narrow emission lines~\citep{2008NewAR..52..227T}, and are further classified according to their broad band SED properties, namely the peak frequency of their Synchrotron emission
~\citep{Padovani1995,FermiSEDs}, from low-energy (radio) peaked to high-energy (X-ray) peaked sources~\citep{1995A&AS..109..267G}. 

Recently, a gamma-ray-emitting blazar, TXS 0506+056, has been identified as the probable astrophysical counterpart to a TeV neutrino emission event~\citep{TXSneutrino}, and since then blazars have been increasingly referred to as the likely sources of the growing population of VHE IceCube neutrinos \citep[e.g.,][]{2018ApJ...865..124M,2018MNRAS.480..192P,Giommidissecting}, putting them right at the center of the emerging multi-messenger %(MM) 
astrophysics.

Numerically dominant in the high-energy catalogues, when identified on the basis of gamma-ray observations their classification is particularly challenging, and about a fourth of the sources in the 4FGL catalogue remain without firmly established counterparts at other wavelengths. This is due to the poor source localisation accuracy (up to several arcminutes for Fermi-LAT), which yields dozens of potential counterparts and renders association ambiguous. The establishment of firm multi-wavelength identifications is nevertheless essential, given the importance of high-energy data for the understanding of the blazar phenomenon and studies of the extragalactic background light ~\citep{2013APh....43..215S}. 

Traditionally, multi-wavelength associations are done mostly on the basis of the angular separation between candidate counterparts from different bands~\citep{4FGLAGN}. Likewise, source classification is usually done manually, cross-matching such multi-wavelength information ~\citep[e.g.,][]{2013ApJS..206...12D,2018A&A...616A..20A}. Recently, many works have emerged that attempt to address the problem of source classification and association through the application of machine-learning algorithms. 

The majority follows a classification strategy based on the gamma-ray data alone, aimed either at identifying AGNs within the sample of unidentified Fermi-LAT sources (e.g.,
~\cite{2020A&C....3200387X} and~\cite{2013MNRAS.428..220H}) or to discriminate between BL Lacs and FSRQs among the population of Blazar candidates of uncertain-type (BCU), for example, in the latest 4FGL (\cite{2019MNRAS.490.4770K} and ~\cite{2019ApJ...887..134K}). Some authors have extended the machine learning approach to the direct problem of multi-wavelength associations, taking into account certain properties of the candidate counterparts such as colours and spectral information to yield more physically-motivated results (e.g.,~\cite{2020ApJS..248...23D} and~\cite{2019ApJ...887...18K}), demonstrating the importance of a cross-band source-identification approach which includes broadband spectral data.

The main goal of this work is the identification of blazars based on non-contemporaneous multi-wavelength information, by using a Deep Learning-based tool. In particular, rather than focusing on providing new source associations, we wish to evaluate the more general reliability of an ML-based classification of blazars within a heterogeneous multi-wavelength database of AGNs. Our work differs from multi-band approaches such as by~\cite{2020ApJS..248...23D} in the fact that we use the complete, heterogeneous, SED information for classification, rather than an approach based on some pre-selected parameterisation of the population. It also innovates with respect to a recent proposal by
~\cite{2020arXiv200503536A}, in the sense that we use a purpose-built deep-learning algorithm rather than more general machine-learning tools. 

We designed a convolutional long-short term memory neural network for the specific problem of distinguishing blazars from a general AGN database based on their complete SED information. We apply minimum or no pre-treatment on the input data, as we aim the algorithm for flexible and direct application to general datasets. This kind of model has presented a high performance in several sequence classification and regression problems, and it is highly adaptable. As the model is defined in a data-driven fashion based on real data, it does not require previous modeling that could lead to systematic errors, although it is still susceptible to possible systematic errors and biases related to data acquisition. 

\par The dataset used in this work is the most extensive yet applied to similar efforts based on previously compiled Blazar and general AGN catalogs, with the data collection made possible by the use of the Open Universe VOU-blazars tool \citep{voublazars}, which retrieves data using Virtual Observatory (VO) protocols from over 65 multi-wavelength catalogues, surveys and spectral databases distributed worldwide, processes them into homogeneous, Galaxy de-absorbed, $\nu F_{\nu}$ fluxes, and builds the SED.

\par Our results show that the neural network was able to separate blazars form other AGNs successfully and to a satisfying degree. A cross-validation procedure was performed and the metrics of a blind test sample show minimal variance between the folds. 

The paper is divided in the following way: Section 2 presents the dataset used in the study; Sections 3 and 4 describe in detail the purpose-built deep-learning algorithm developed, as well as the training of the neural network; Section 5 presents the results of the deep-learning classification; Section 6 discusses physically the results and their general applicability; and finally, Section 7 summarises the discussion and presents some future perspectives.

\section{Classification Dataset}
\label{sec:data}

Sample selection and definition is a fundamental step in machine learning applications. In an effort to develop the best possible learning algorithm capable of identifying and correctly classifying the SED of blazars out of a more general population of Active Galaxies, we sought to build the largest set of well-populated SEDs. 
To this end, we used the master list of blazars compiled by the Open Universe for blazars program, which includes over 6,000 objects. This list has been built combining the ROMA-BZCAT \citep{bzcat}, featuring 3,561 spectroscopically confirmed blazars of all types, the 3HSP catalogue, which includes 2,013 blazars of the HSP type \citep{Chang2019}, and the list of newly discovered blazars in the Fermi 4FGL-DR2 catalog \citep{4FGL,4FGLDR2}. The sample of non-blazar AGN was selected from the  million quasar catalog V6.4  \citep[milliquas][]{Flesch2015}, accepting sources that are also included in several multi-frequency catalogs and are not included in the open universe master list of blazars, containing mostly bright quasars and Seyfert galaxies, with a total 8,467 sources. The complete sample is then comprised of 14,555 objects.

\par 
Collecting broadband multi-frequency data for such a large number of sources is notoriously difficult. To perform this task, we used the VOU-Blazars \citep{voublazars}, a VO-based tool developed within the context of the United Nations Open Universe Initiative (UNOUI; see e.g.~\citet{ou4blazars}) and available through the Open Universe portal currently maintained by the Italian Space Agency (ASI)\footnote{www.openuniverse.asi.it}. The tool aims to facilitate large-scale data discovery and aggregation by searching and collecting all spectral information available online in open access, VO-compatible databases, for a given source -- not only blazars, despite the name. The data collection also includes information available from the ASI Space Science Data Center (SSDC) and Open Universe-generated products, the latter entirely available in VO-format.

The data selection for construction of the SEDs was done with VOU-Blazars using a 15 arcsecond cone search from the reference positions of the sources in the sample above. We finally obtained 6,045 blazar and 7,962 non-blazar SEDs, combining for a total of 14,007 objects in our final database. The remaining objects from our base sample had no data points available for retrieval through the VOU-Blazar tool. This means that the final dataset comprises the most complete (and probably nearly exhaustive) VO-based SED sample publicly available online. No further downselection of the sources was made in order to make the sample more homogeneous. We consider this heterogeneity an important aspect of our dataset, since we aim at a classification algorithm that is as comprehensive as possible. 

\begin{figure}
    \centering
    \includegraphics[width=\linewidth]{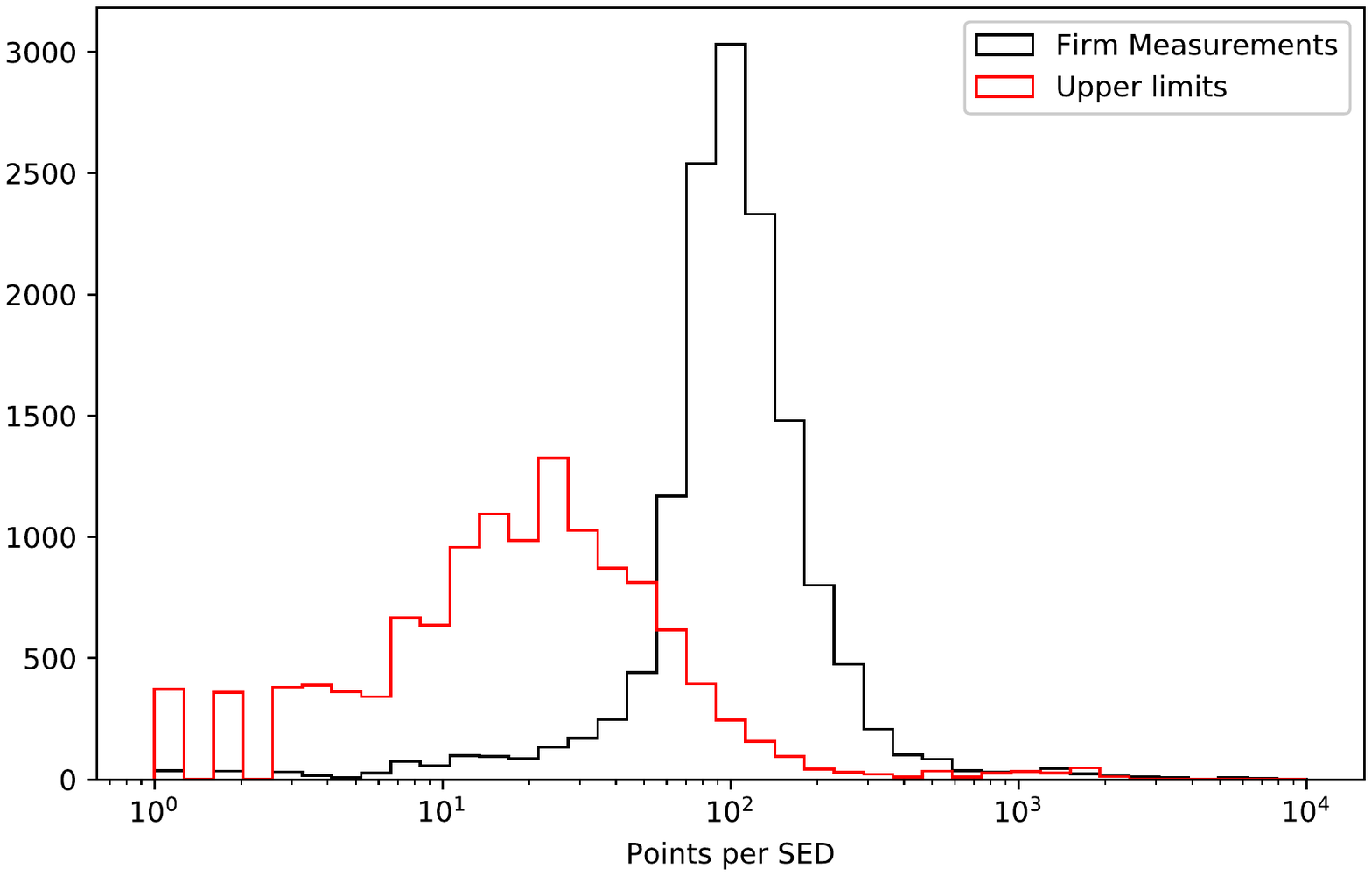}
    \caption{Histogram of the number of sources per amount of data points in the SED. Most objects in our sample have SEDs composed of a few hundreds of flux measurements.}
    \label{fig:histo_points}
\end{figure}

While some of our SEDs are very complete, with a large amount of data points over the entire spectral range (the best studied objects have thousands of observational points throughout the electromagnetic spectrum, and spanning a wide range of flux levels), for some sources only very few, non-contemporaneous measurements are available. The majority of the SEDs collected is composed of a few hundreds of data points, as shown in Fig.~\ref{fig:histo_points}. Resilience in classifying the SEDs even in such "data starved" conditions is another goal of our algorithm, and later in the paper we will analyse in detail this resilience as we move to poorly sampled cases, as well as the influence of the presence of non-contemporaneous measurements in the SED classification.

\begin{figure*}
\centering
    \subfloat[Non-Blazars.]{\includegraphics[width=.5\textwidth]{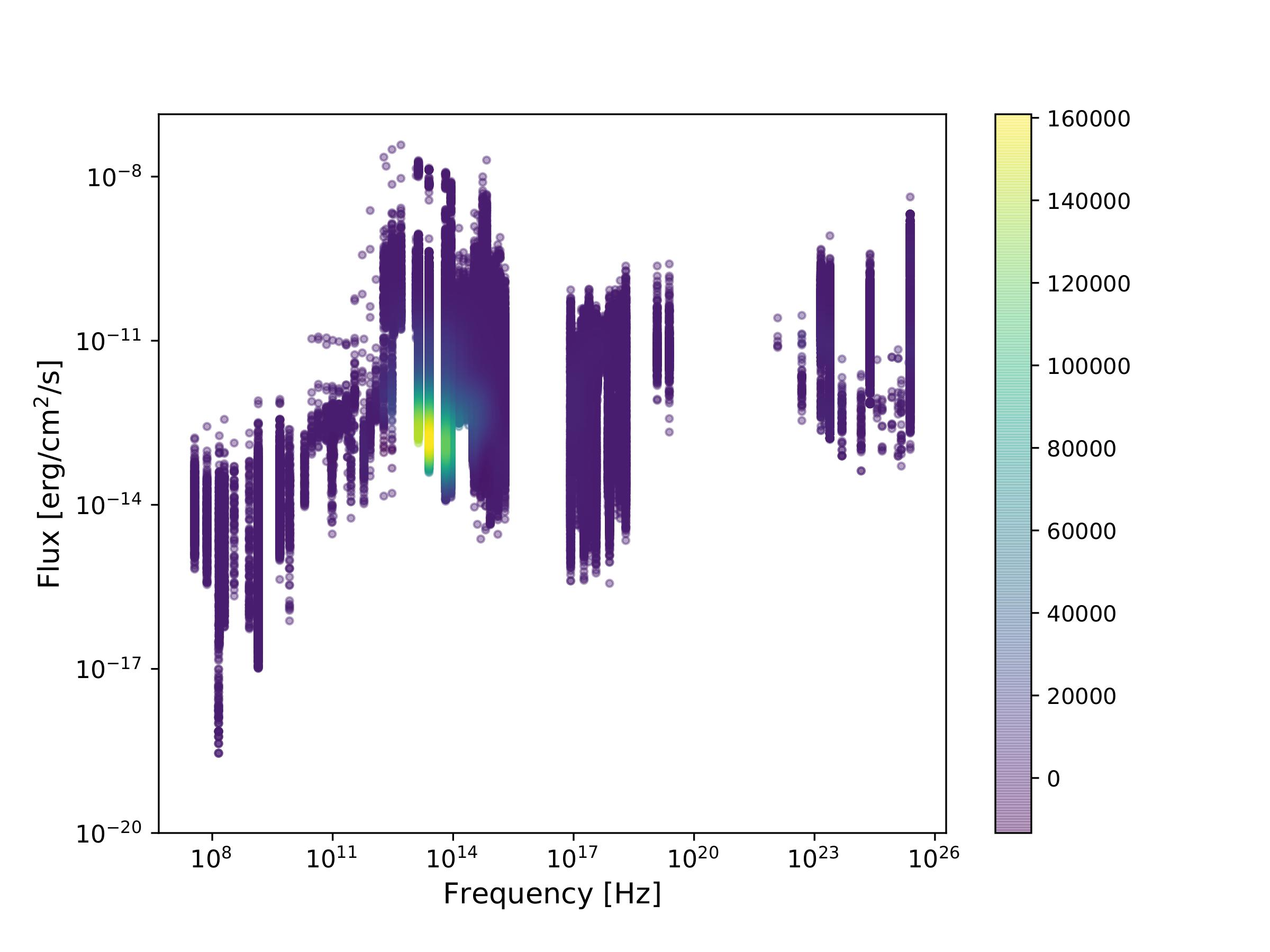}}
    \hfill
    \subfloat[Blazars]{\includegraphics[width=.5\textwidth]{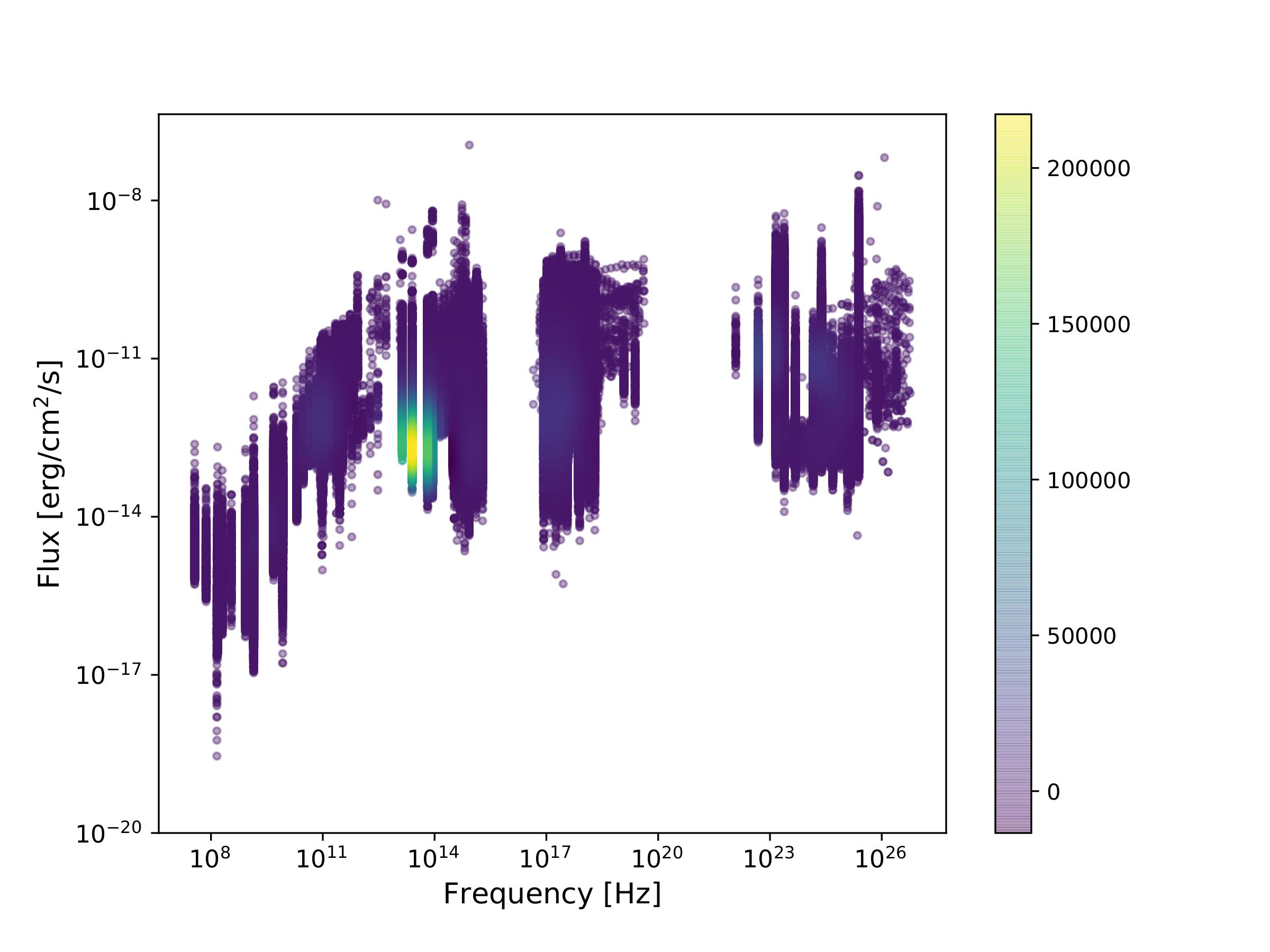}}
\caption{Density plots of the SEDs in our dataset, separated between (a) non-blazars and (b) blazar. Lighter colors indicate higher density of data. Observe the difference in spectral coverage at the highest energies between the two populations of sources. The two figures are shown to scale and one can visually perceive some systematic differences between the SEDs of the two populations, specially in the X-ray region.}
\label{fig:seds}
\end{figure*}

Fig.~\ref{fig:seds} collects all the SED data for the blazar and non-blazar samples used in our analysis. To give an idea of the amount of data generally available at each spectral band, the number of points is indicated with a colour scale. Only flux measurements are indicated in the figure, and flux upper limits have been flagged out whenever present and excluded from the classification analysis. As can be seen, the sub-mm to near-IR is the best sampled spectral range, with nearly all SEDs having some coverage in these bands. Whereas the gamma-ray coverage is fairly non-uniform between the two classes, towards the lower energies there is virtually no difference in the spectral coverage between blazars and non-blazars.

One can visually perceive some systematic difference between the SEDs of the two populations of sources, the key element we want to capture with our neural network on an individual source basis. The difference in spectral coverage at the highest energies between the two populations of sources is also readily noticeable, with blazars being much better sampled in $\gamma$-rays than non-blazars sources, as expected. In total, our dataset assembles over 1,900,000 flux measurements and almost 600,000 upper limits. 

\begin{figure}
   \centering %height=\textheight
   \includegraphics[width=0.5\textwidth,keepaspectratio]{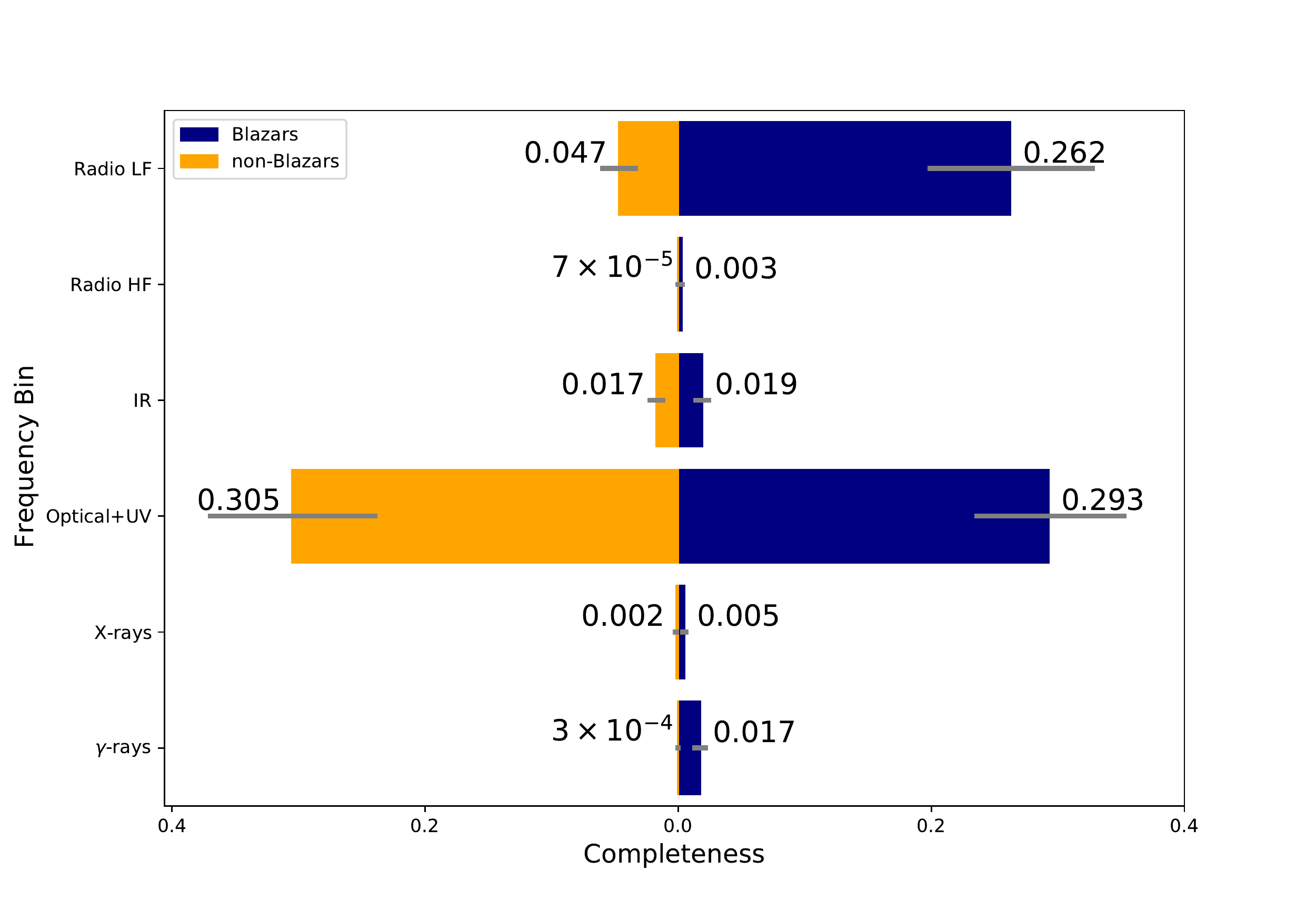}
      \caption{Completeness of the SED sampling per bin of frequency, for Blazars (blue) and non-Blazars (orange). The bars indicate the mean completeness for that frequency bin, with one standard deviation denoted by a grey bar. The bins are defined in Table
     \ref{tab:bands}.}
     \label{fig:coverage}
\end{figure}

A complimentary diagnosis of the dataset is provided by Fig.~\ref{fig:coverage}, which shows the completeness of the SED sampling per frequency bin, i.e., the ratio of sources with at least one firm observation in a frequency to the total, averaged over all frequencies in a given bin. The Infrared (IR) bin shows little difference between Blazars and non-Blazars, as well as the Optical/Ultraviolet (UV) and and X-ray bins. On the other hand, the Blazar completeness in Radio, both low frequency (LF) and high frequency (HF), is significantly larger than non-Blazars one; the former are radio-loud, while there is a significant amount of radio-quiet AGNs among the latter. The vast majority of the sources in this dataset observed in the $\gamma$-ray region are Blazars, as expected by the these objects' extreme properties  mentioned in Section~\ref{sec:introduction}. 

\begin{table}
    \centering
    \caption{Definition of the frequency bins from Fig.~\ref{fig:coverage}}
    \label{tab:bands}
    \begin{tabular}{|c|c|}
    \hline 
    Bin Name    & Frequency Range (Hz) \\
    \hline 
    Radio LF   & $<1\times10^{10}$ \\
    Radio HF  &  $<3\times 10^{11}$ \\
    IR  &  $<4\times10^{14}$ \\
    Optical+UV &  $<3\times10^{16}$ \\
    X-rays  &  $<2.4\times10^{20}$ \\
    $\gamma$-rays  &  $>2.4\times10^{20}$ \\
    \hline
    \end{tabular}
\end{table}

The heterogeneity of the dataset makes it challenging for automated classification methods, and it is a good test for deep learning algorithms to handle a dataset like this.

\section{Deep Learning Classification}
\label{sec:deeplearningmodels}

Deep Neural Networks \citep[henceforth DNNs;][]{Goodfellow-et-al-2016,lecun2015deep} constitute a particular case of artificial neural networks that are distinctively useful in many structured data such as images and time series. The basic structure of any Neural Network is a mathematical representation of a neuron $y_j$, which consists of a linear combination of inputs $x_i$ commonly associated with a transfer function $\varphi$ (named activation function):
\begin{equation}
y_j =  \varphi \left( \sum_{i=0}^m w_{ij} x_i + b\right),  
\label{eq:nn_activation}
\end{equation}
where $w_{ij}$ are the weights and $b$ the bias, which are the parameters to be optimized in a DNN for a certain dataset. The neurons are arranged in layers that determine how two sets of neurons are connected. Therefore, the inputs $x_i$ can be either the initial input data or the data already processed by the previous neuron layer.

Convolutional Neural Networks (CNNs) are particular layers inspired in pattern recognition tasks developed in brains of mammals \citep{hubel1962receptive,lecun2015deep}. They consist of kernels that are convolved with the data as they flow through the DNN. For a given Kernel $K$ and array $x(i)$ we may define the convolution operation as:

\begin{equation}
s_j(i) =  \left( \sum_{i=0}^m K(i-m) x(i) \right),    
\end{equation}

The deep model typically contains several stacked layers with different functions, such as dimensionality reduction (MaxPooling), neuron activation functions, and fully connected (or dense) layers of neurons.

DNN and CNN-based models have emerged as the State-of-The-Art in several computer vision tasks, such as image classification \citep[see, e.g.][]{ILSVRC15,zhang2018multi,lindeep2018,john2015pedestrian,peralta2018use}, Natural Language Inference\citep{zhang2019semantics} and pose estimations, as well as applications to a range of multidisciplinary fields, for instance in medicine \citep{shi2020review,norgeot2019call} and Oil \& Gas reservoir characterization \citep{dias2020automatic,valentin2018estimation,valentin2019deep}. 
Their performance in classification tasks is reported to overcome humans for some datasets \citep{challenge}.
Traditional Neural Networks and DNNs have demonstrated to be very effective at extracting the internal patterns of data in astrophysical classification tasks. For instance, in morphological classification of galaxies \citep{dieleman2015rotation}, in the identification of Strong Lensing systems \citep{bom2017neural,challenge,2018MNRAS.473.3895L}, star and galaxy separation \citep{kim2016star}. Recently, this has also been extended to inverse modelling of astrophysical systems such as Strong Lensing \citep{bom2019deep,2017Natur.548..555H,pearson2019use,levasseur2017uncertainties,morningstar2018analyzing}. 

\subsection{Long-Short Memory units}
\begin{figure}
   \centering
   \includegraphics[width=0.48\textwidth]{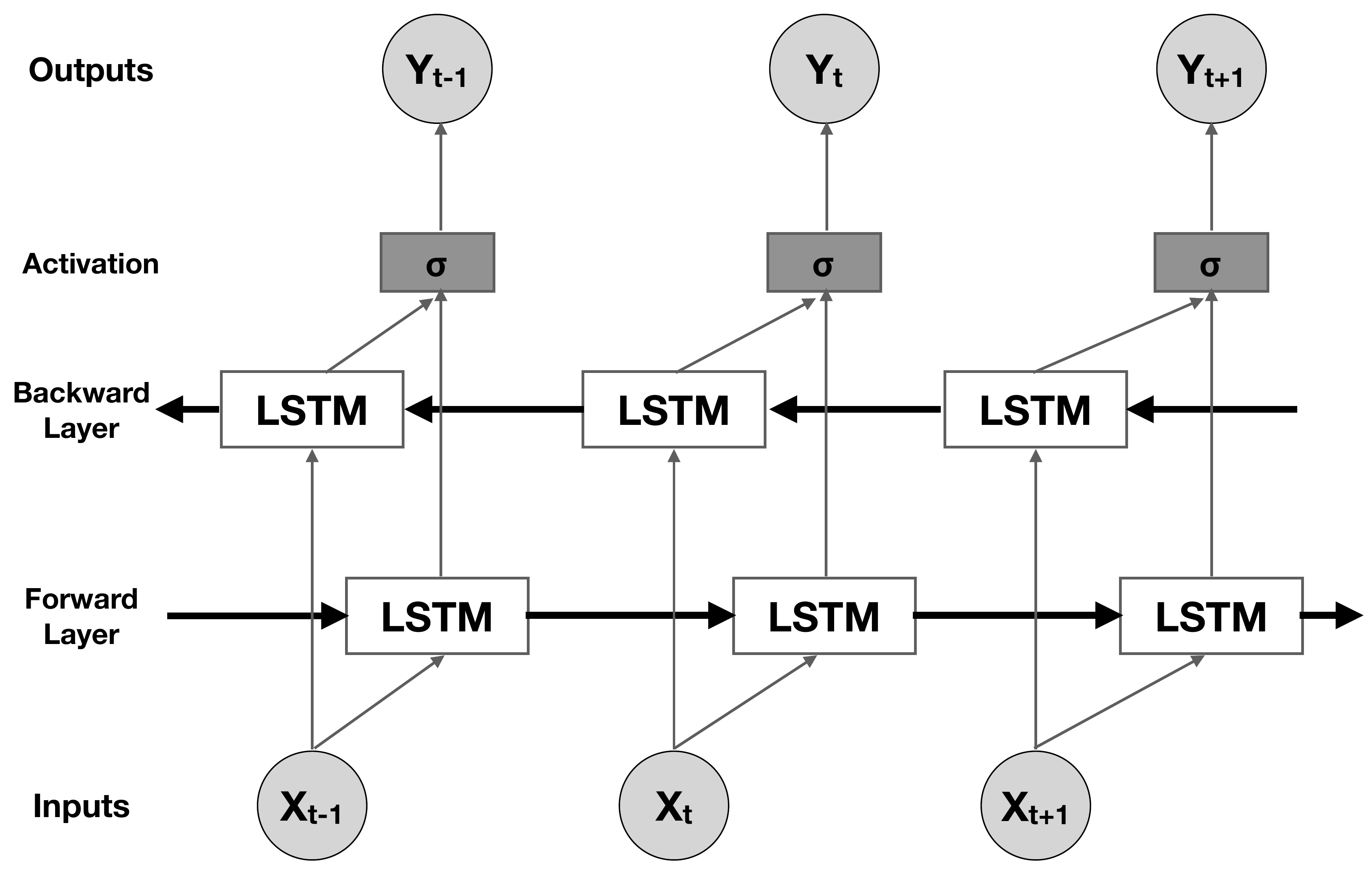}
      \caption{Overall scheme of a Bidirectional LSTM. Information coming from the previous points ($x_{t-1}$, $x_{t-2}$, ...) and the next points ($x_{t+1}$, $x_{t+2}$, ...) is used to produce the output $Y_t$ at each point $x_t$.}
     \label{fig:bilstm_concept}
   \end{figure}

Many of the pattern recognition tasks in computer vision problems are related to sequence or time series classification. To address this task, one needs to develop a DNN capable of analyzing sequential data, such as times series, language translation or curve classification, where data points present strong correlation to its neighbours. One such DNN model is a Recurrent Neural Network \citep[RNNs; see, e.g., ][]{schuster1997bidirectional,medsker1999recurrent,pascanu2013difficulty} and neural networks with Long Short Term Memory Units (LSTMs) are among the most popular types of RNN. These techniques are suited to process sequential information through a series of iterations of the dataset. Thus this kind of approach is widely used for text and audio processing \citep[see, e.g.,][]{sutskever2011generating, graves2013speech}. 
LSTMs were also successfully applied to astronomical data problems, for instance, in the spectral classification of Supernovae and light curve classification \citep{muthukrishna2019dash,pasquet2019pelican}. The main concept behind LSTMs is defining a NN architecture that can keep the relevant information from previous points while effectively removing unnecessary data. These operations are performed by the so-called remember and forget gates, respectively. These are learnable weights connected to different activation functions that are able to correctly balance the information that must be kept or removed. However, there are cases in which the relevant information can be both in the previous points or the next ones. This is the case where we are interested in the general form of the curve. To address this kind of data, a common choice is the bidirectional LSTM \citep{schuster1997bidirectional}, so that for each point of the dataset, there are two LSTM working together analyzing the data in both directions. Figure \ref{fig:bilstm_concept} shows a simplified representation of how a bidirectional LSTM works: at each point, the output ($Y_t$ in Figure \ref{fig:bilstm_concept}) is created by combining an LSTM carrying information from the next points (Backward Layer) with another carrying information from previous points (Forward Layer).

\subsection{Model Architecture}

One of the defining characteristics of a Blazar SED is its shape with two bumps. Since any SED is composed of ordered pairs, we can view the SED of a Blazar as an ordered sequence in which each point is influenced by points coming before (lower energy) and after (higher energy), so that the overall shape is maintained even if the frequency and amplitude of the peaks change. Thus, one natural path for our architecture is to use a Bidirectional Recurrent Network in order to preserve information from "the future" (higher energies) and "the past" (lower energies). In this way, the network will use information coming from both higher and lower frequencies at each point to determine the overall shape of the SED.
\par
\begin{figure}
   \centering
   \includegraphics[width=0.4\textwidth,keepaspectratio]{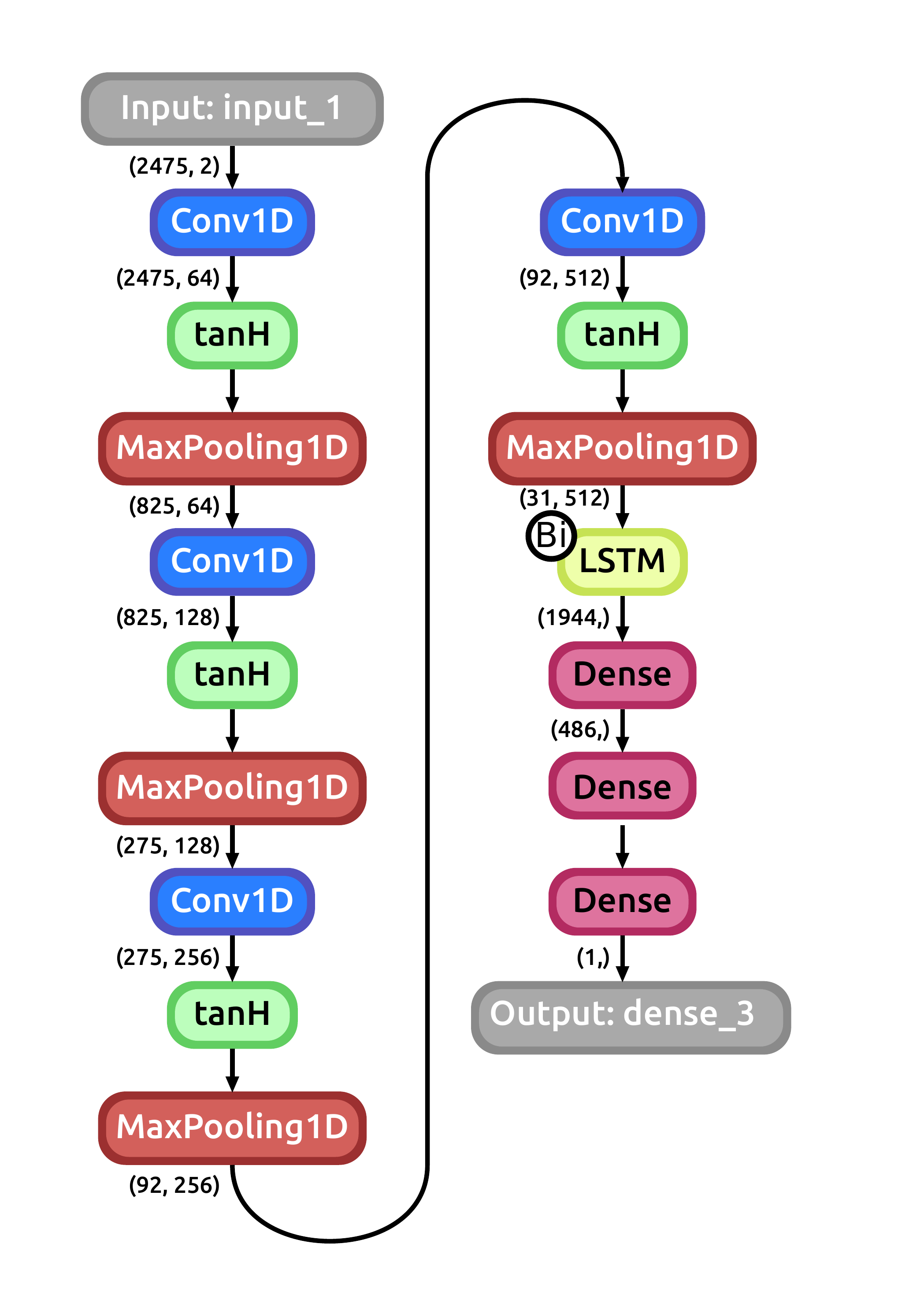}
      \caption{The architecture of our Neural Network. The input data goes through each layer sequentially, starting from the input layer and being transformed according to each layer configuration until the desired output: a single number representing the probability that an SED is from a blazar. The numbers in parentheses show the shape of the data whenever a layer changes it.}
         \label{fig:bilstm}
   \end{figure}

 Our model was built using \textit{Keras}\footnote{https://keras.io} with a \textit{TensorFlow}\footnote{https://www.tensorflow.org} backend in Python, and its architecture is shown in Fig. \ref{fig:bilstm}. The first four blocks, composed of a 1D convolutional layer with a kernel size of 3 followed by a tanh activation and a 1D Max Pooling layer with size 2, are designed to reduce dimensionality and capture the relevant features of the SED. The convolutional layers have an increasing number of filters in each block so that the net can probe finer and finer scales by generating more feature vectors. Afterwards, the output is fed into a Bidirectional LSTM layer before we map the output to the predicted variable through a series of fully connected layers. The net has a total of 16 layers and a little over nine million parameters. The final output is a number corresponding to the probability that a given SED is a blazar. 
\label{sec:models}

% --------------------------------------------------------------------
% SubSection: Training
% --------------------------------------------------------------------
\section{Training the Deep Learning Model}
\label{sec:training}

The input to the machine learning algorithm consisted of a vector of flux measurements and a flag for upper limits for each source in the sample. The data was ordered in frequency to reproduce the natural sequence of the SED. Since the frequency sampling was different for each source, the missing data were numerically flagged to maintain the same fixed size for the flux input vectors. In total, the SED was split into 2745 frequency points, which were the same for all input vectors. 

Whenever multiple flux measurements were available for a given frequency, only one value was chosen as input to the neural network. Here we chose to systematically select the highest flux points. This choice was made aiming to mitigate a potential mixture of source states (quiescent and flaring) that is likely present in the sample of non-contemporaneous measurements. The choice of the highest fluxes also naturally emphasises the jet's spectral signatures, which is the most active and dominant component of the SED in blazars. Other sources of thermal emission that could be present, such as the host galaxy or accretion disk, tend to be swamped by the jet during active states, evidencing the SED's double-humped shape. This selection of the data lead to the best success rate in the classification, compared to no flux selection, as either the flux variability or the larger discrepancy in the number of flux points between well and ill-sampled SEDs seemed to confuse the network. That said,  choosing the average or even the minimum flux level have little to no impact on the final results, since for the vast majority of the sources the flux range is not significantly large. The metrics (see Section \ref{sec:results}) are compatible within one standard deviation with our original approach, using the same cross-validation procedure described below. This data selection step is, therefore, not considered critical to the application of the algorithm. Furthermore, to assess if the flux errors could affect the final result, we trained the model replacing each flux point by a random variable taken from a gaussian distribution, with mean equal to the flux value and standard deviation equal to the flux error. The metrics were compatible with our results within one standard deviation, showing that not including the errors does not impact the final results.

Finally, given the relative paucity of $\gamma$-ray data and the very inhomogeneous coverage difference between blazars and non-blazars in this band (see Figs.~\ref{fig:seds} and \ref{fig:coverage}), the classification algorithm was run in two different sets: the complete sample using the entire SED information, and a reduced sample excluding all frequencies in the $\gamma$-ray region ($\log\nu > 20$). This reduced the number of frequencies input to the network to 2391, all within the Synchrotron region of the spectrum. All fluxes were re-scaled to fall in the range $[0,1]$ before being fed to the neural net. 

One crucial step before feeding the data to the net is to divide the sample into training, validation, and testing subsets. Training subsamples are used by the backpropagation algorithm to update the internal weights and biases of each layer of the model in order to minimise the loss function. The validation set is used after each training step to evaluate the network's performance and prevent overfitting. Finally, the testing set is used only after the training has finished producing the classification results. The network uses neither the validation nor the testing sets to update its weights, and since the testing data is only used after training is completed, we can have an idea how the model will generalise to yet unseen data. 

A common method to try to avoid biases due to using a small part of the data for testing is to apply cross-validation: the data is split into $k$ different subsets, where $k-1$ are used for training and validation, and one for testing. By repeating the procedure $k$ times using a different subset at each iteration for testing, we ensure that the results were obtained using the whole dataset and not just part of it. This procedure is called \textit{K-fold} cross-validation, with $K$ the number of subsets (folds) the data is split. 

\begin{figure*}
\centering
    \subfloat[]{\includegraphics[width=0.49\linewidth]{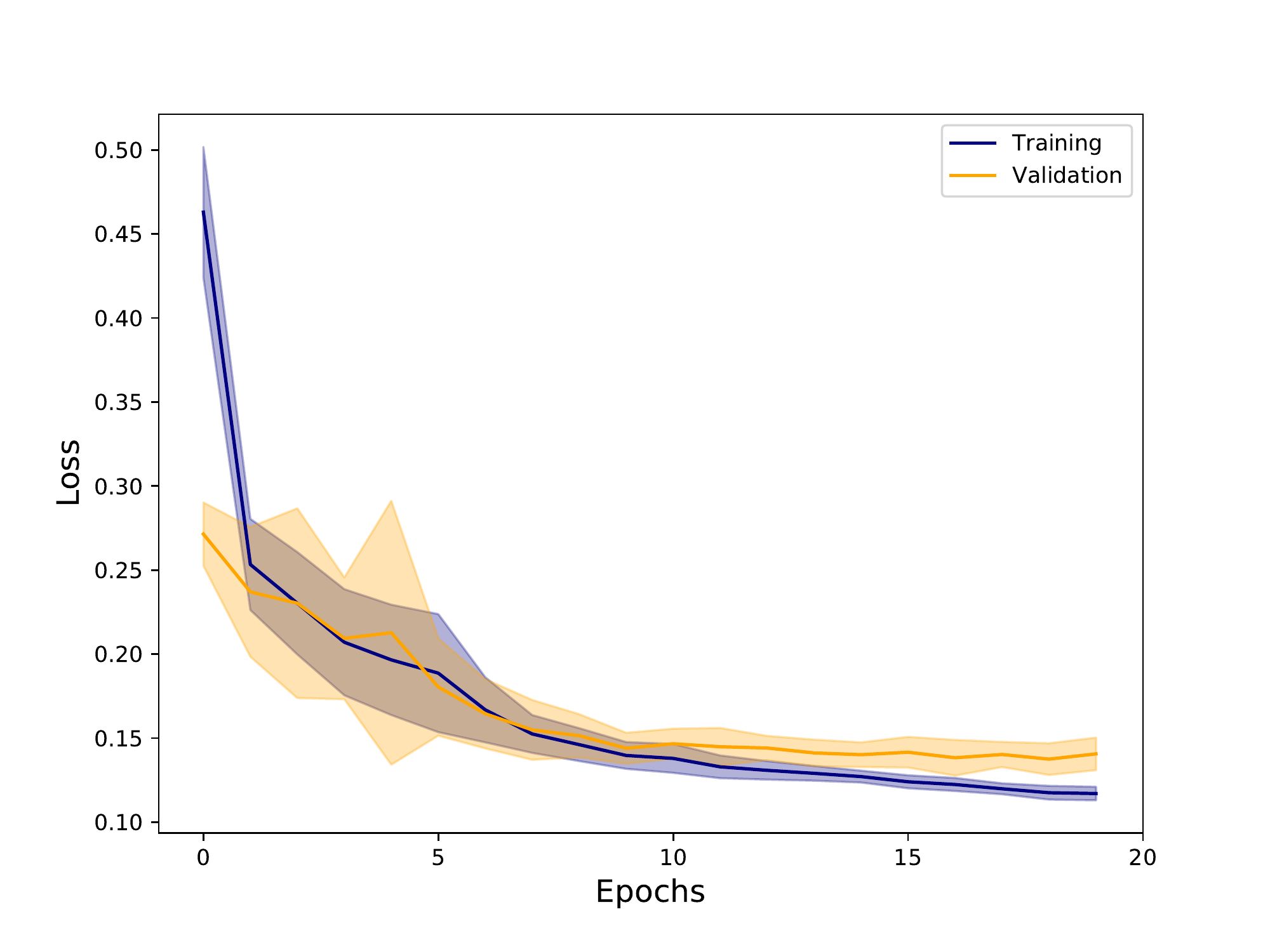}}
    \hfil
    \subfloat[]{\includegraphics[width=0.49\linewidth]{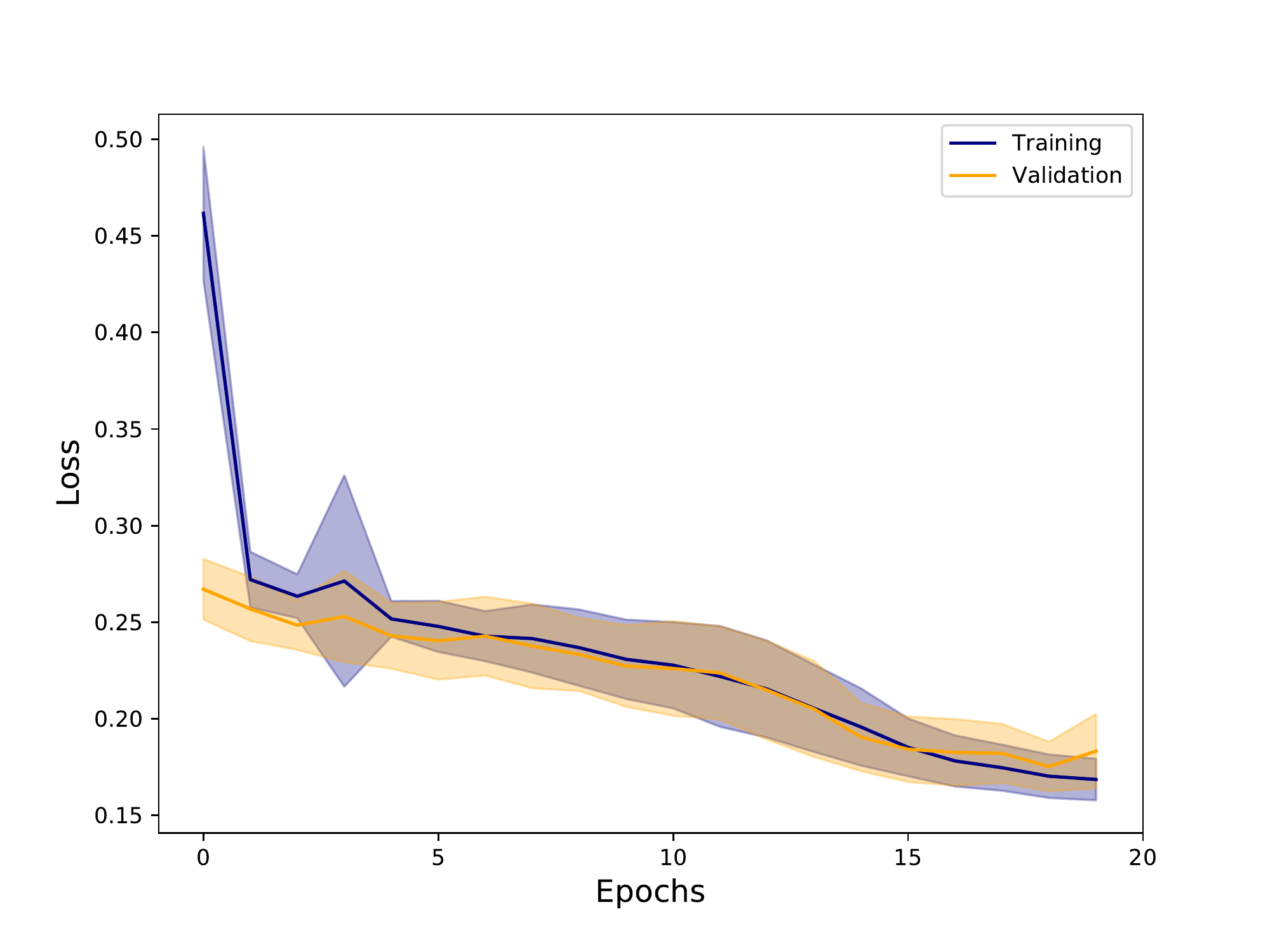}} \\
    \subfloat[]{\includegraphics[width=0.49\linewidth]{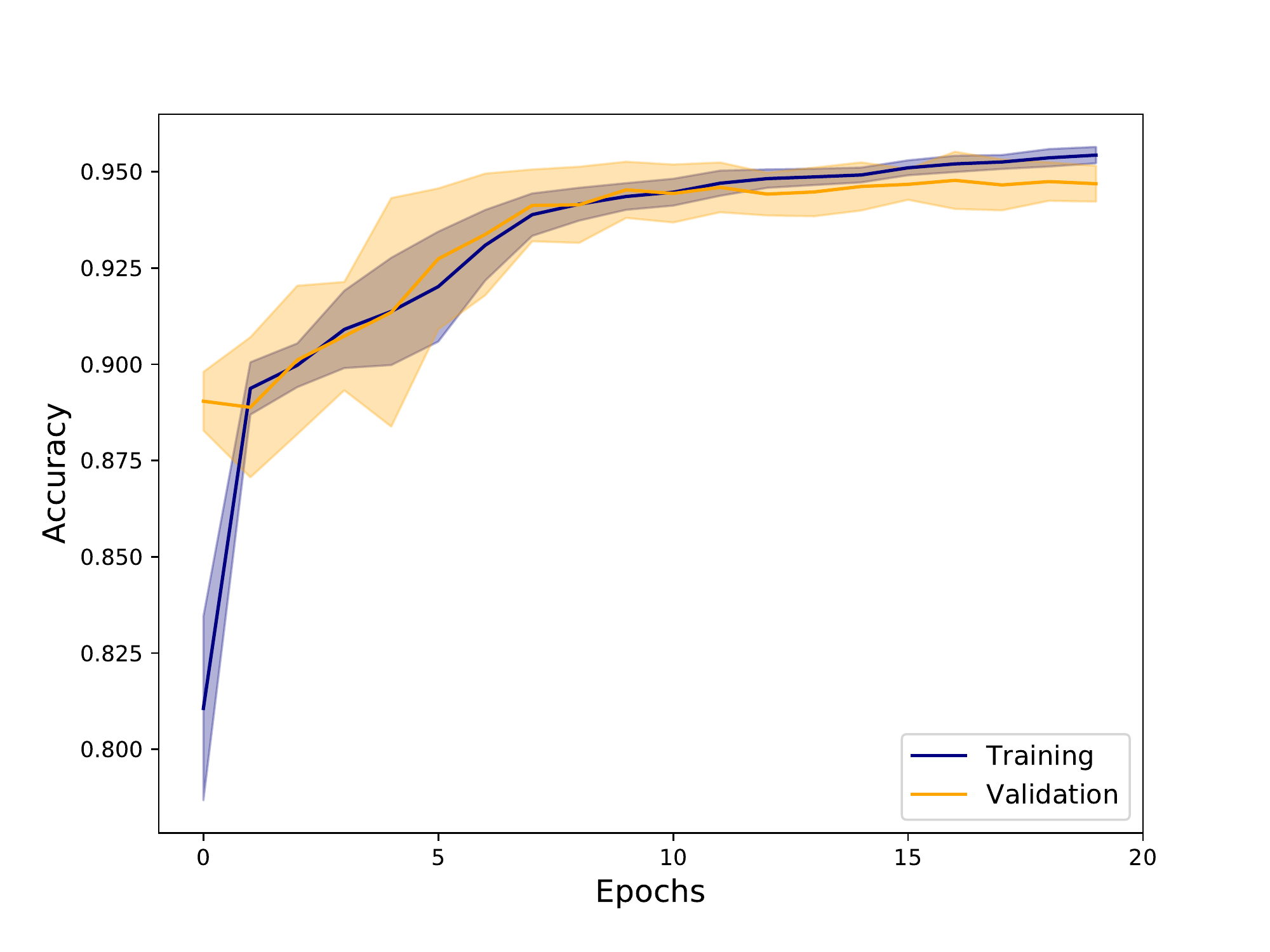}}
    \subfloat[]{\includegraphics[width=0.49\linewidth]{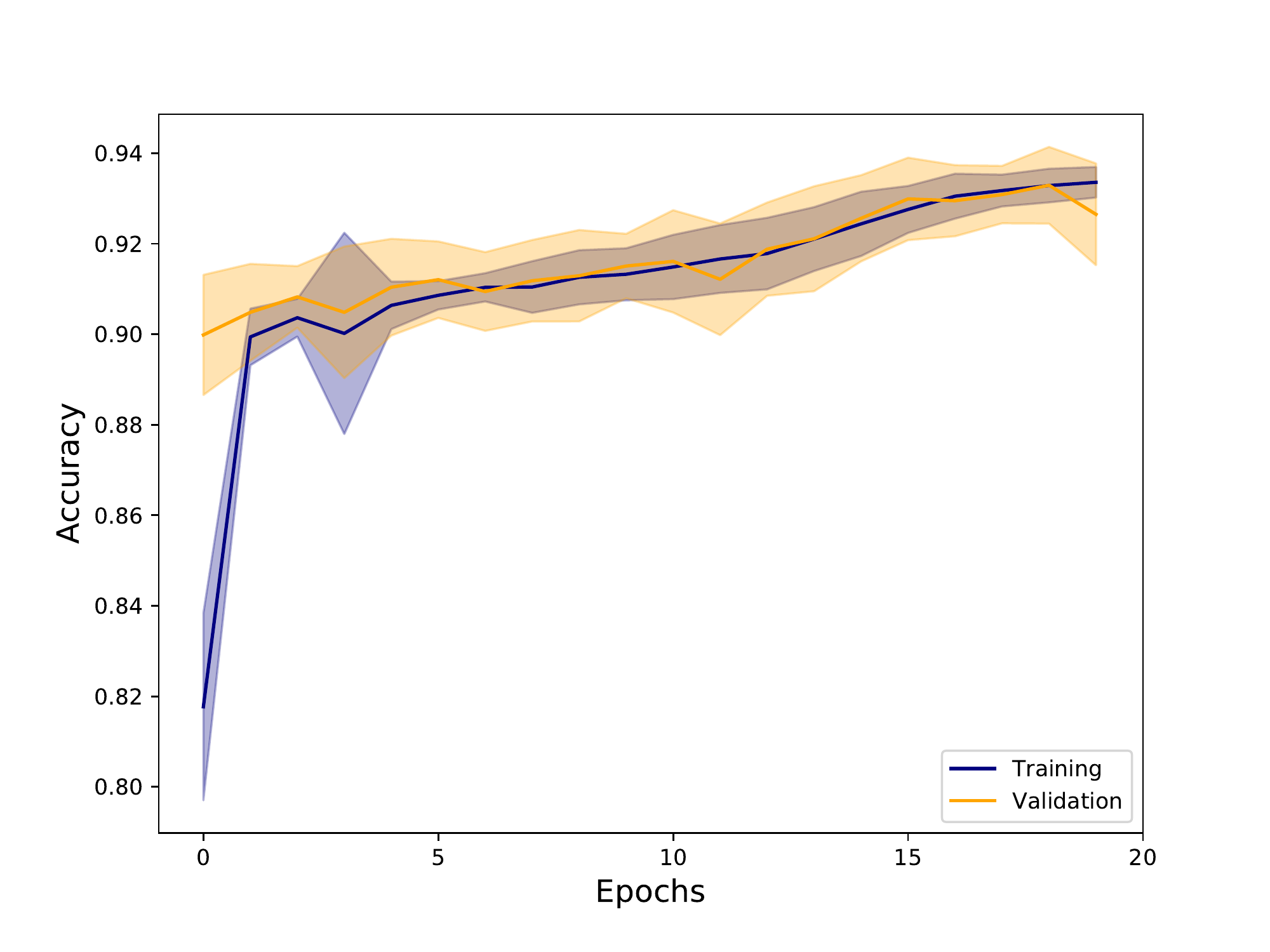}}
    \caption{Mean loss and accuracy for training (blue) and validation (orange) sets, with $\gamma$-ray data (a, c) and without (b, d). The shaded regions correspond to one standard deviation from the mean.}
    \label{fig:training}
\end{figure*}

We used a 10-fold cross-validation in our analysis so that, at each fold, $10\%$ of the data is used for testing and the rest is split $90/10$ between training and validation. The total number of samples in each subset, per fold, is shown in Table~\ref{tab:foldings}. We trained the model for 20 epochs for each fold, using a batch size of 128 and the Adam optimizer \citep{adamopt}. The training time on a GeForce 1080 was $\approx 25$ minutes/20 epochs for the reduced sample and $\approx 30$ minutes/20 epochs for the full sample. The mean accuracy and loss for the training and validation subsamples are shown in Fig.~\ref{fig:training} for the reduced and the complete sample, with the shaded region corresponding to one standard deviation. 

\begin{table}
    \caption{Distribution of SEDs for each fold.}
    \centering
    \begin{tabular}{cccc}
    \hline
    Total samples & Train per fold & Validation per fold & Test per fold \\
    \hline
    $14007$ & $11345\, (81\%)$ & $1261\, (9\%)$ & $1401\, (10\%)$ \\
    \hline
    \end{tabular}
    \label{tab:foldings}
\end{table}

\section{Results}
\label{sec:results}
Here we present the results of the classification for the testing sub-sample as described in Section \ref{sec:training}. One of the main tools to analyse the performance of an ML classifier is the Receiver Operating Characteristic (ROC) curve. A ROC curve shows the true positive ratio (TPR) and false positive ratio (FPR) at different separation thresholds between the classes. They are defined as

\begin{equation}
    \mbox{TPR} = \frac{TP}{TP + FN}, \quad\quad \mbox{FPR} = \frac{FP}{FP + TN},
\end{equation}

\noindent where TP are the true positives (correctly predicted as the positive class), FP the false positives (incorrectly predicted as the positive class), TN the true negatives (correctly predicted as the negative class), and FN the false negatives (incorrectly predicted as the negative class). In this work we consider blazars to be the positive class, but this choice is immaterial and the opposite could have been chosen with the same results. 

The Area under the ROC Curve (AUC) measures the degree of separability of a given model: the higher the AUC, the better the model is at  differentiating between the classes. A perfect classifier would have an AUC of 1 since it does not make any wrong predictions and therefore $FPR=0$, always. This means that a good classifier will have the TPR increase rapidly while keeping the FPR low, making the ROC curve very steep near the origin. 

\begin{figure*}
   \centering 
   \subfloat[With $\gamma$-ray data.]{\includegraphics[width=.49\textwidth,keepaspectratio]{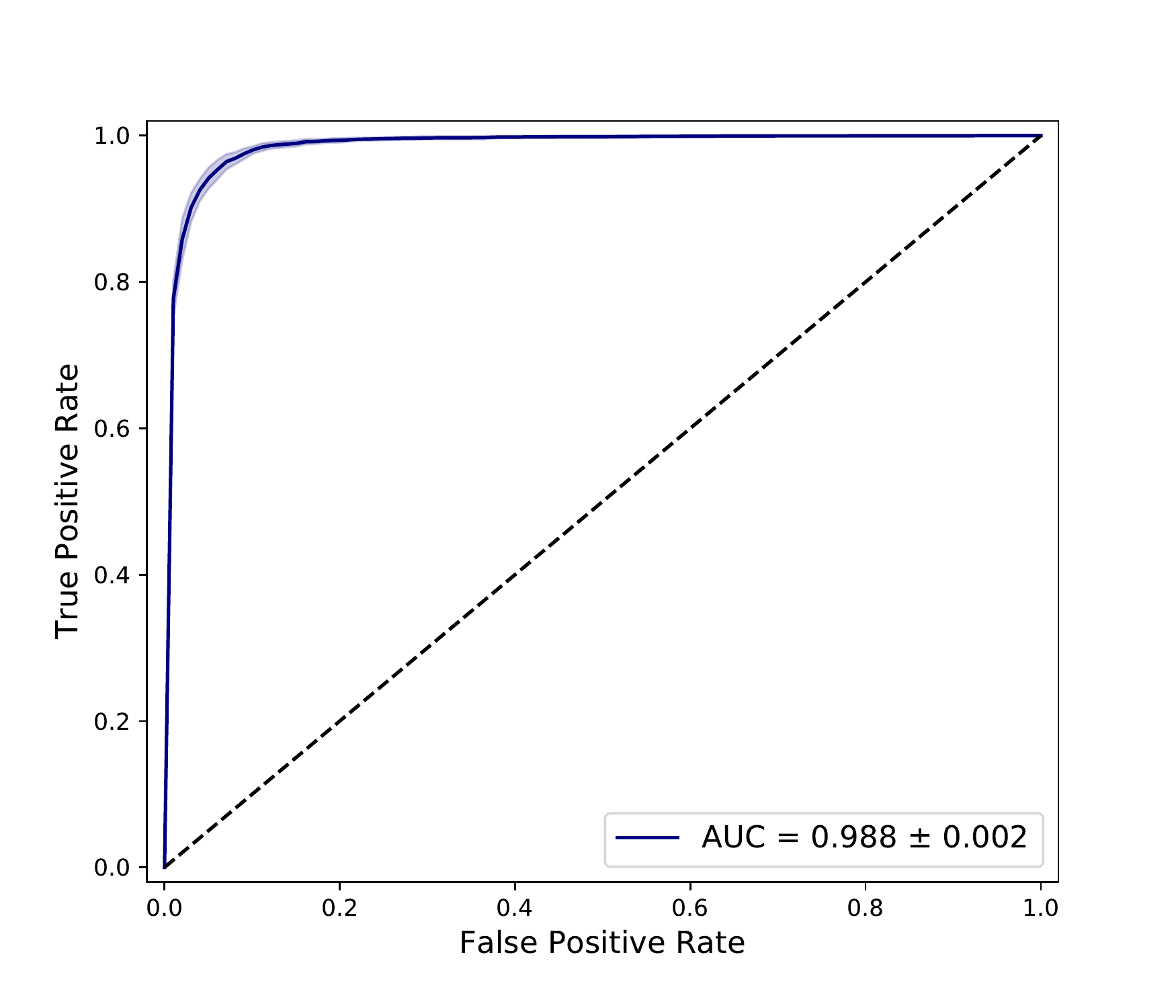}}
     \hfill
   \subfloat[No $\gamma$-ray data.]{\includegraphics[width=.49\textwidth,keepaspectratio]{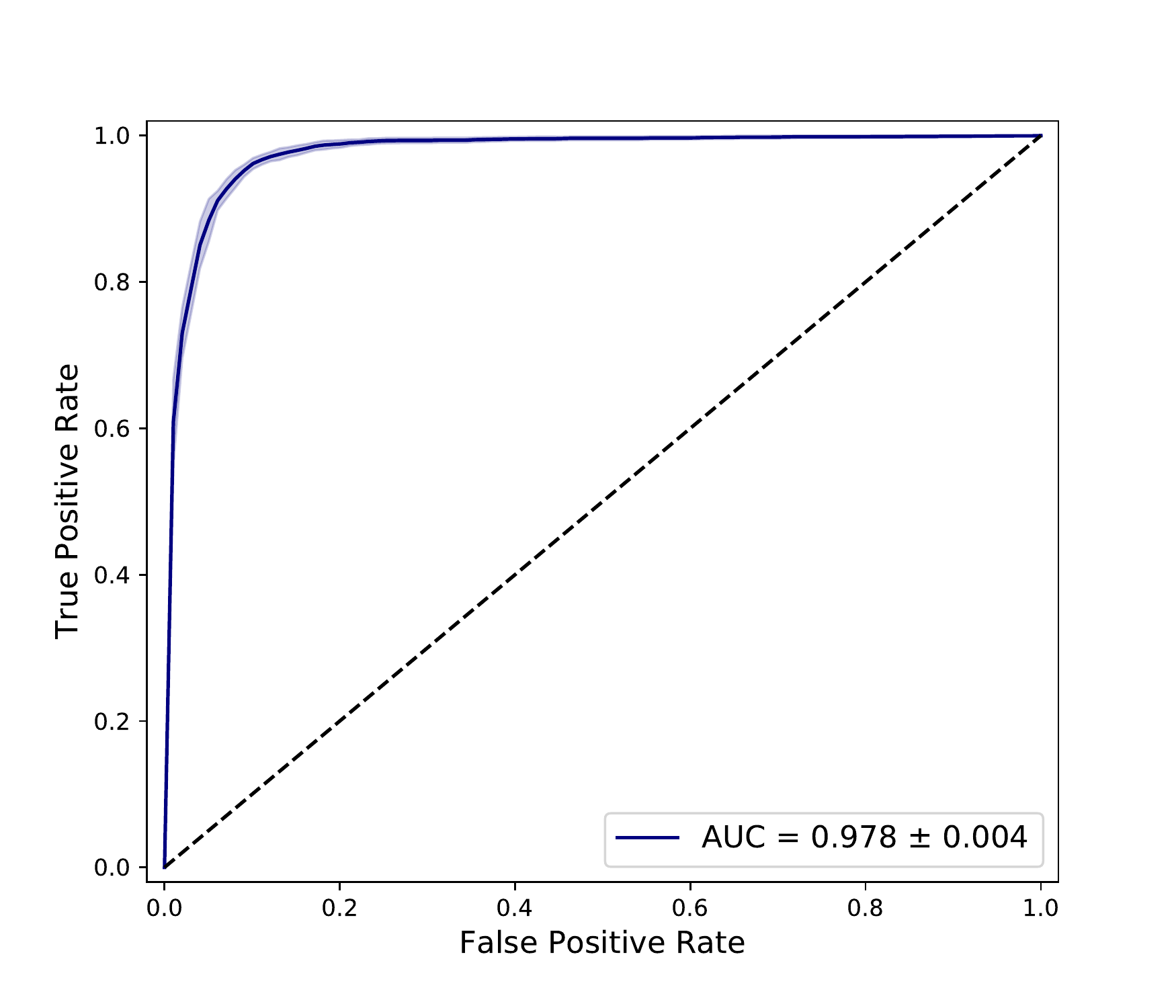}}
    \caption{Mean ROC curves for Blazars (blue) and non-Blazars (red). The transparent band indicates one standard deviation and the black dashed line indicates a random classifier.}
\label{fig:rocs}
\end{figure*}

\par The average ROC curves for the sets with and without $\gamma$-ray data are shown in Fig.~\ref{fig:rocs}. The net had a similar performance in both cases and was able to classify the sources with similar results based only on the low energy part of the SED, even though the amount of data is decreased by more than $10\%$. This lack of degradation in the network's capability in the absence of the gamma-ray data is likely due to the correlation between the Synchrotron and Inverse-Compton bumps of the SED. In either case, the net had a remarkably good result in the test sets with minimal variance between folds, showing the robustness of the model when used with unseen data.

\par Another tool to analyse the performance of the model is the Precision-Recall (PR) curve, where Recall is the same as TPR and

\begin{equation}
    \mbox{Precision} = \frac{TP}{TP + FP}.
\end{equation}

Intuitively, \textit{Precision} can be thought of as the ability of the net to not misclassify a non-blazar (blazar) as a blazar (non-blazar), while Recall is the ability of the net to identify all blazar (non-blazar) samples. Like the ROC, the PR curve is built taking the precision and recall values at different thresholds of separation between classes. Here, however, there is a trade-off between them: if we change the threshold to increase Recall by lowering the number of false negatives, this will automatically increase the number of false positives, decreasing Precision. The AUC for a PR curve reflects the average Precision for different Recall values, and, as for the ROC, the higher the better.

\begin{figure*}
   \centering 
   \subfloat[With $\gamma$-ray data.]{\includegraphics[width=.49\textwidth,keepaspectratio]{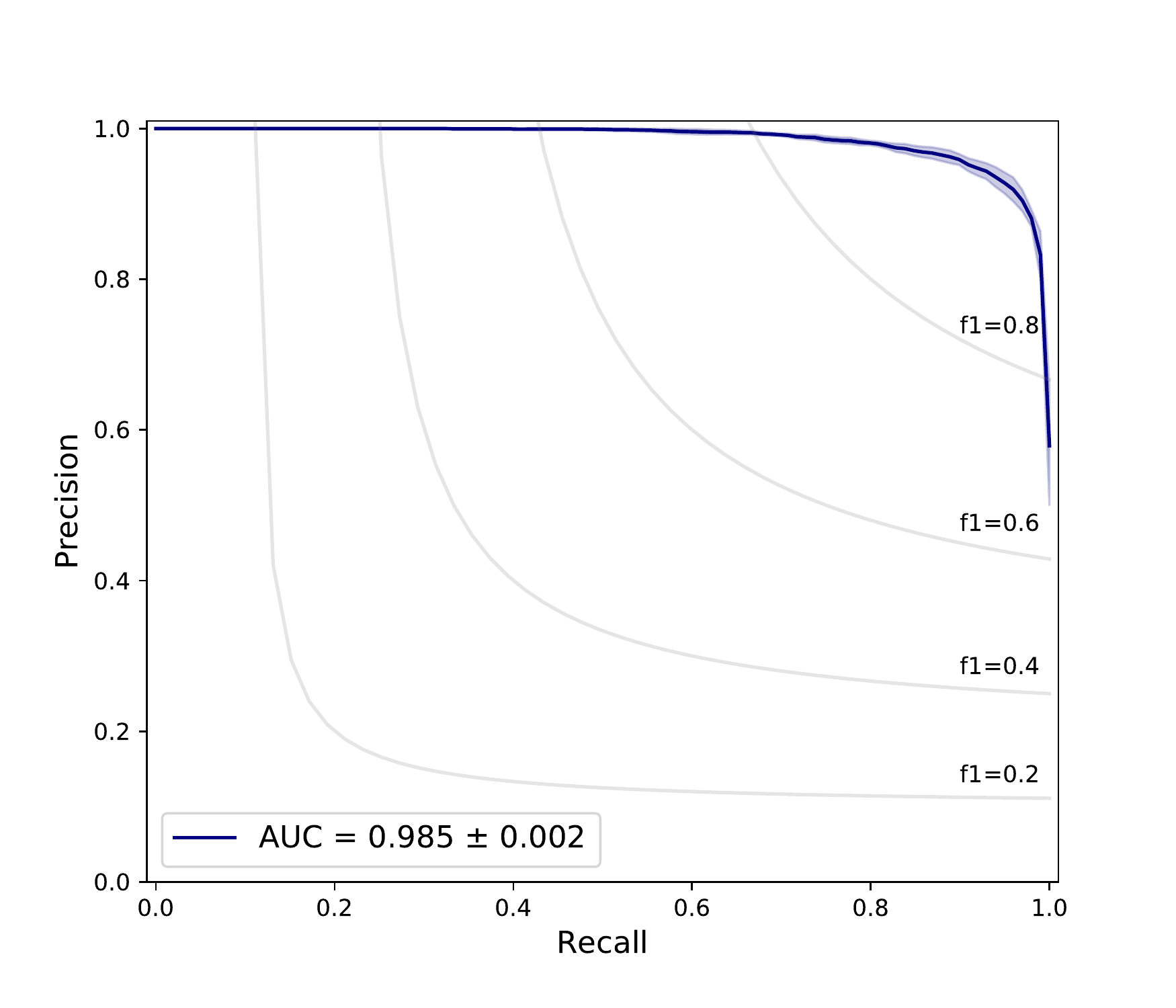}}
    \hfill
   \subfloat[No $\gamma$-ray data.]{\includegraphics[width=.49\textwidth,keepaspectratio]{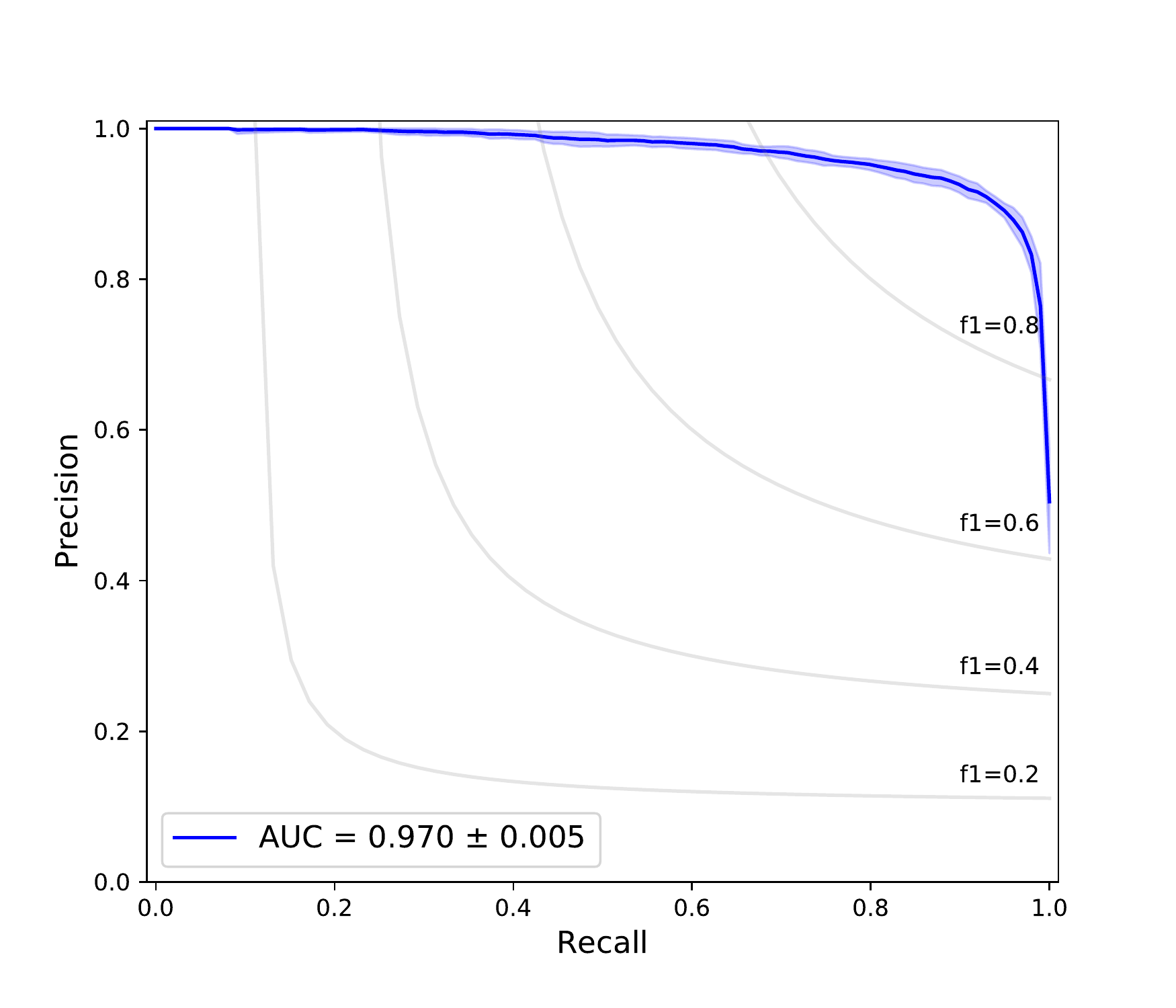}}
         
    \caption{Mean Precision Recall curves for Blazars (blue) and non-Blazars (red). The transparent bands indicate one standard deviation from the mean and constant F1 scores are indicated by grey lines. }
    \label{fig:PR}
\end{figure*}

\par The average PR curve for both sets of data is shown in Fig.~\ref{fig:PR}, together with lines of constant F1 score (the harmonic mean between Precision and Recall). The net's performance is similar in both cases just like the ROCs, with satisfying results and, again, minimal variance between the folds.

\begin{table}
\caption{Confusion matrix for the 10-fold cross-validation: (top) with $\gamma$-ray data; (bottom) without $\gamma$-ray data.}
\centering
\begin{tabular}{|l|l|l|}\hline
\backslashbox{True Class}{Predicted} & Non-Blazar & Blazar \\ \hline \hline
Non-Blazar & 94.4\% $\pm$  0.9\% & 5.6\% $\pm$  0.9\% \\ \hline
Blazar & 4.6\% $\pm$  0.8\% & 95.4\% $\pm$  0.8\% \\ \hline
\end{tabular}
\newline
\vspace*{0.3 cm}
\newline
\begin{tabular}{|l|l|l|}\hline
\backslashbox{True Class}{Predicted} & Non-Blazar & Blazar \\ \hline \hline
Non-Blazar & 92.5\% $\pm$  0.7\% & 7.5\% $\pm$  0.7\% \\ \hline
Blazar & 6.0\% $\pm$  0.6\% & 94.0\% $\pm$  0.6\% \\ \hline
\end{tabular}
\label{tab:confusion_matrix}
\end{table}

As a final metric to evaluate the model's performance, we report the mean confusion matrix for the sets with and without high energy data in Table~\ref{tab:confusion_matrix}. Here the horizontal labels correspond to the true classes while the vertical labels to the model predictions. To determine the class predicted as the output of the net, it is first necessary to specify a threshold of separation between the classes. We chose the optimal value as the one corresponding to the point in the ROC curve closest to the point $(0,1)$, i.e., the threshold at which the model's ROC curve would be as close as possible to a ROC curve of a perfect classifier (since it would be always right, the FPR is zero while the TPR is one, always). 

The confusion matrices here show the percentage of blazars/non-blazars that the model predicted as each of the classes. For example, considering the set with $\gamma$-ray data, our model correctly predicted $94.4\%$ of the non-blazars, while $5.6\%$ of them were misclassified as blazars. In the set without $\gamma$-ray data, the net correctly predicted $92.5\%$ of the non-blazars and misclassified $7.5\%$ of them. 
\par The results above show that the net's performance was excellent in both datasets, with minimal difference between them. Given that $\gamma$-ray data is scarcer and usually more challenging to obtain, a tool that can reliably identify blazars using only the low energy part of SEDs would be extremely useful for catalog building and blazar searches in general.

The misclassified sources (false positives and false negatives) could provide some clues of the model's limitations, possibly translating these to physical properties. However, a visual analysis of those SEDs at each fold did not show anything singular about them. The relative number of false positives and negatives at each fold is small, so this analysis might not be significant.
  %%%%%%%%%%%%%%%%%%%%%%%%%%%%%%%%%%%%
  %%=================================+%%%%%%
  %%%%%%%%%%%%%%%%%%%%%%%%%%%%%%%%%%%%%%%%%%%%%%%%%%
\section{Gradient Analysis}
The results shown in Section \ref{sec:results} show that our network was able to successfully classify blazars from other AGNs. On a basic level, neural networks work by trying to find patterns in a multi-dimensional space. Given that the classification was successful, one might try to extract at least part of these patterns, possibly helping in understanding the physics behind Blazars. 
\par One way of extracting physical meaning from the classification is to analyse which features were most important to the network when classifying a source as a Blazar or non-Blazar. This can be done by calculating the gradients of the final layer in a trained net w.r.t. the inputs: larger values mean that the weights will be changed the most during backpropagation, and, as can be seen from Eq. \ref{eq:nn_activation}, neurons with larger weights will impact the final result more. Thus, the larger the absolute gradient value of a feature, more importance was given to it by the net when performing the classification. Since for each SED we feed to the net the order of frequencies is the same, we can use these gradients to map the importance of each frequency, trying to tie this to a more physical understanding.  
\par
\begin{figure*}
   \centering
   \subfloat[Blazars.]{\includegraphics[width=.49\textwidth,keepaspectratio]{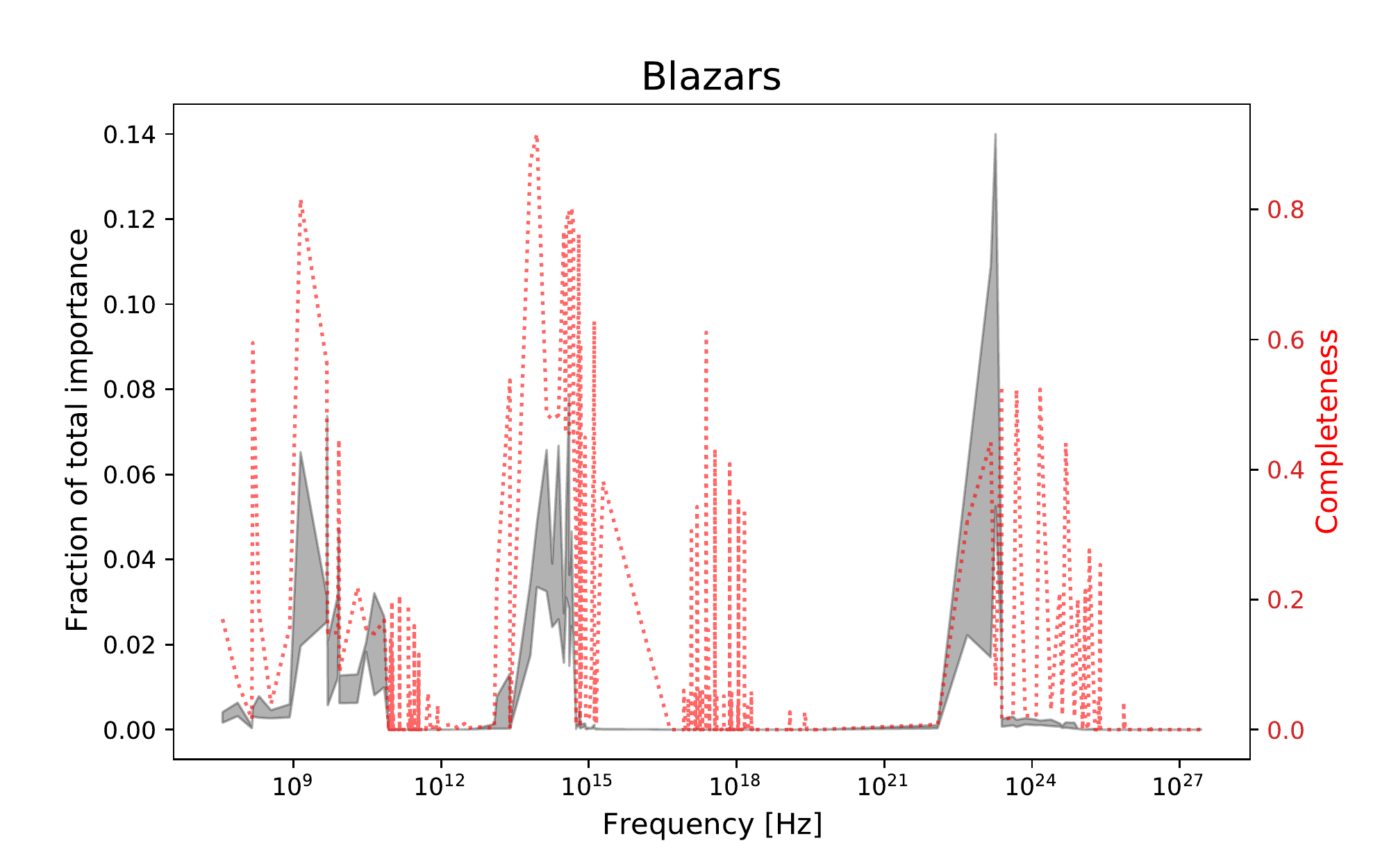}}
    \hfill
   \subfloat[Non-Blazars.]{\includegraphics[width=.49\textwidth,keepaspectratio]{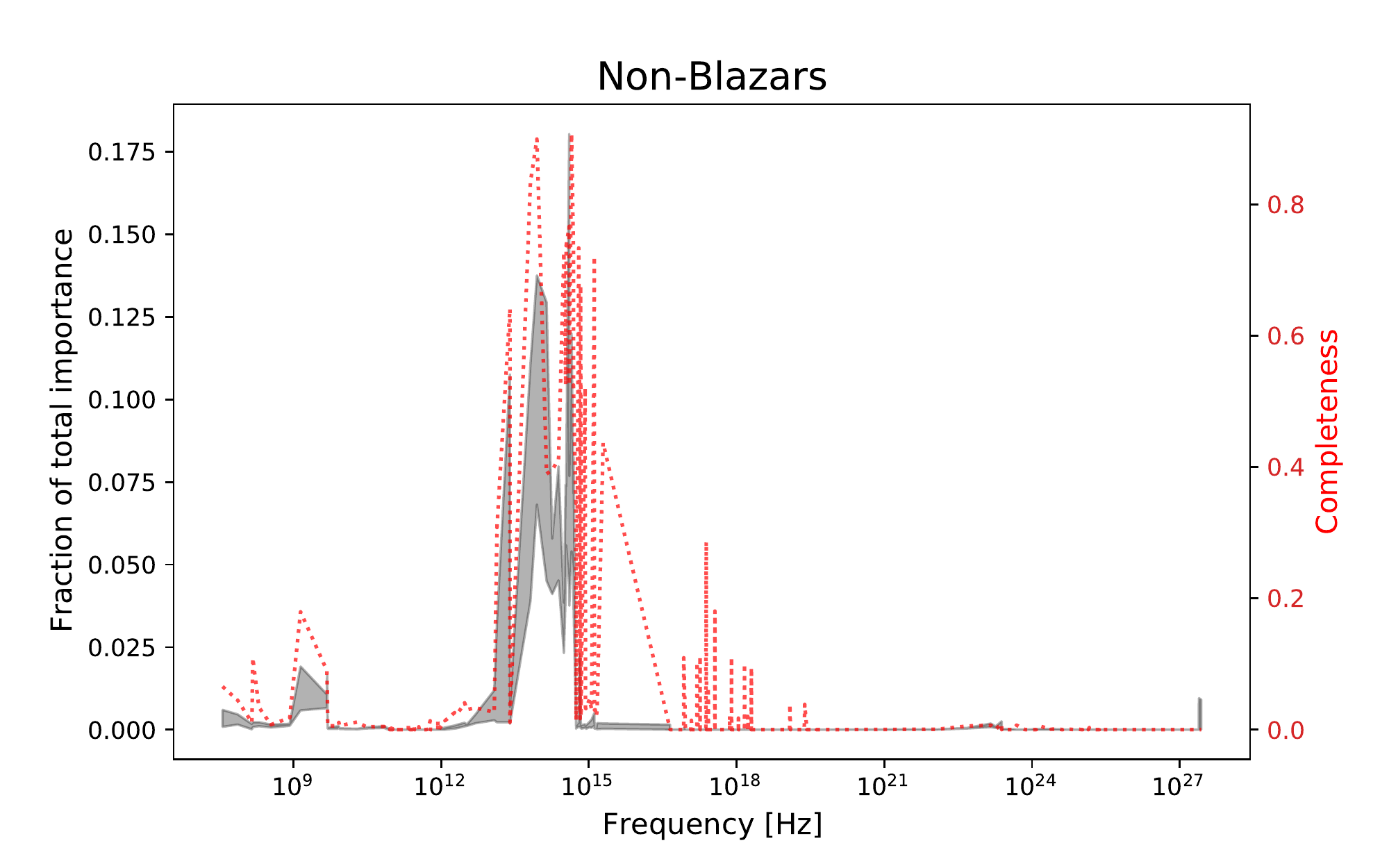}}
         
    \caption{Fraction of the total importance for the four best performing nets in the cross-validation (black) and the completeness of the dataset (red), per frequency.}
    \label{fig:attention}
\end{figure*}

 The gradients were calculated using the four nets from the K-fold cross-validation which achieved the lowest validation losses to check for the robustness of the results. The relative importance given by the net to each frequency is calculated through the gradients: for each SED, we normalize the gradients, which are then summed up to get the ratio of total importance given by the net to each frequency. Fig.~\ref{fig:attention} shows this ratio and the completeness, at each frequency, for both Blazars and non-Blazars. We have defined completeness as the fraction of sources with at least one firm observation (not taking upper limits) at any given frequency.

\par  One important aspect is that the regions where the dataset is most complete do not correlate very well with the regions of most importance. This is especially evident in the X-ray region, where there are some frequencies with good completeness but the net ignored them. This indicates that there are other factors at play besides data availability during the net's training; it possibly learned something that we can translate into physical properties of AGNs. 
\par The $\gamma$-ray region was important when classifying blazars but was largely ignored for other AGNs. Blazars are the most dominant source in the extragalactic high-energy sky, and this is reflected in our dataset by the significant difference in the number of observations in this spectral region between the different classes (see Figs.~\ref{fig:seds} and \ref{fig:coverage}), making it extremely likely that a source with $\gamma$-ray data is a Blazar. 
\par Moving to lower energies, we see that the X-ray region was not important, despite having good completeness in several frequencies. Since the synchrotron peak frequency in Blazars can vary by several orders of magnitude, the SED in the X-rays could be increasing, decreasing, or flat, depending on the source's class. In the non-Blazar case, the SED is usually dominated by thermal emission from accretion; however, especially in quiescent states or in sources where the synchrotron peak frequency is at low values, the emission in the X-rays can be thermal even in Blazars. Furthermore, the X-ray region presents very large variability, and even though we systematically selected the highest flux value at each frequency to mitigate this effect, it nonetheless could contribute to the overall confusion between both classes in this region. These effects combined could explain the negligible importance of the X-rays in the classification.

\par The IR region was significantly important for the classification for both classes; in non-blazars (especially radio quiet), the bulk of the infrared emission is from thermal dust, causing a small bump in the SED (the so-called IR bump), while for blazars (and most radio-loud AGNs), the IR is dominated by the non-thermal synchrotron radiation.  \par On the lowest energies, the Radio emission is somewhat important when classifying blazars, and plays a small role for non-blazars. Since our dataset contains a good number of radio quiet objects, it is natural to expect the radio emission to be one of the fundamental characteristics used to identify Blazars; however, our non-Blazar sample still contains a solid number of radio-loud AGNs, so that the classification cannot be done based solely on the radio emission. 
\par To try and map the gradients to physical properties of the emission, we looked at the 20 most important frequencies for both classes, for all four nets; of those, ten are common to Blazars and non-Blazars, in the Radio, IR, and Optical regions. With these, we tried to construct parameters that would present at least some degree of differentiation between the classes. One of the most common ones to select Blazar SEDs is the slope between different frequencies,
\begin{equation}
    \alpha_{\nu_1\nu_2} = -\frac{\log(F_{\nu_1} / F_{\nu_2})}{\log(\nu_1 / \nu_2)},
\end{equation}
where $F_{\nu}$ is the flux and $\nu$ the frequency. 
\par The simplest way to select radio-loud sources is from the Radio spectrum; these objects have a flatter slope than radio-quiet ones. However, Blazars are not the only radio-loud class, so this simple approach is usually not enough to distinguish Blazars from non-Blazar AGNs. Other approaches use the plane determined by the WISE Infrared satellite colours, wherein $\gamma$-ray emitting Blazars lie in a specific region \citep{wisestrip, wisepca}. This method can be refined further by also using the Radio-IR and IR-X-ray slopes \citep{1whsp, 2whsp}, or the so-called q-parameter \citep{wibrals}, defined as 
$$
q = \log\left(\frac{F_{IR}}{F_{Radio}}\right). 
$$

\begin{table}
    \caption{Ten most important frequencies for the classification and the Telescope/Survey that made the obervation.}
    \centering
    \begin{tabular}{cc}
    \hline
    Frequency (Hz) & Observer \\
    \hline
    $1.4\times 10^9$ & NVSS, FIRST, NORTH20 \\
    $4.8\times 10^9$ & NORTH20 \\
    $8.817\times 10^{13}$ & WISE W1 \\
    $1.389\times 10^{14}$ & 2MASS K$_s$\\
    $1.804\times 10^{14}$ & 2MASS H \\
    $2.427\times 10^{14}$ & 2MASS J \\
    $3.116\times 10^{14}$ & PANSTARRS y \\
    $3.382\times 10^{14}$ & SDSS \\
    $3.462\times 10^{14}$ & PANSTARRS z \\
    $3.987\times 10^{14}$ & PANSTARRS i\\
    \hline
    \end{tabular}
    \label{tab:important_frequencies}
\end{table}
\par
\begin{figure*}
   \centering 
   \subfloat[]{\includegraphics[width=.33\textwidth,keepaspectratio]{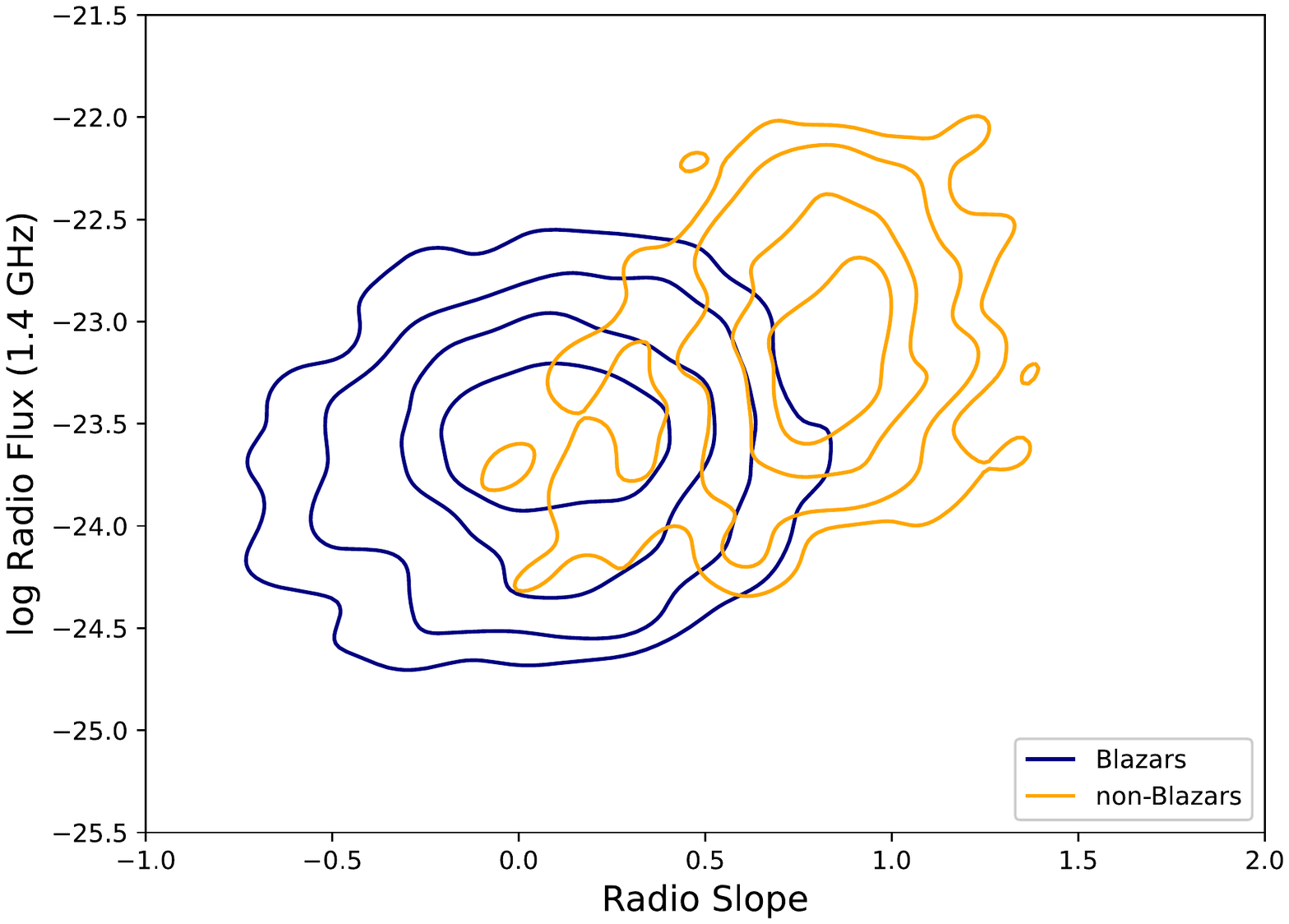}}
    \hfill
   \subfloat[]{\includegraphics[width=.33\textwidth,keepaspectratio]{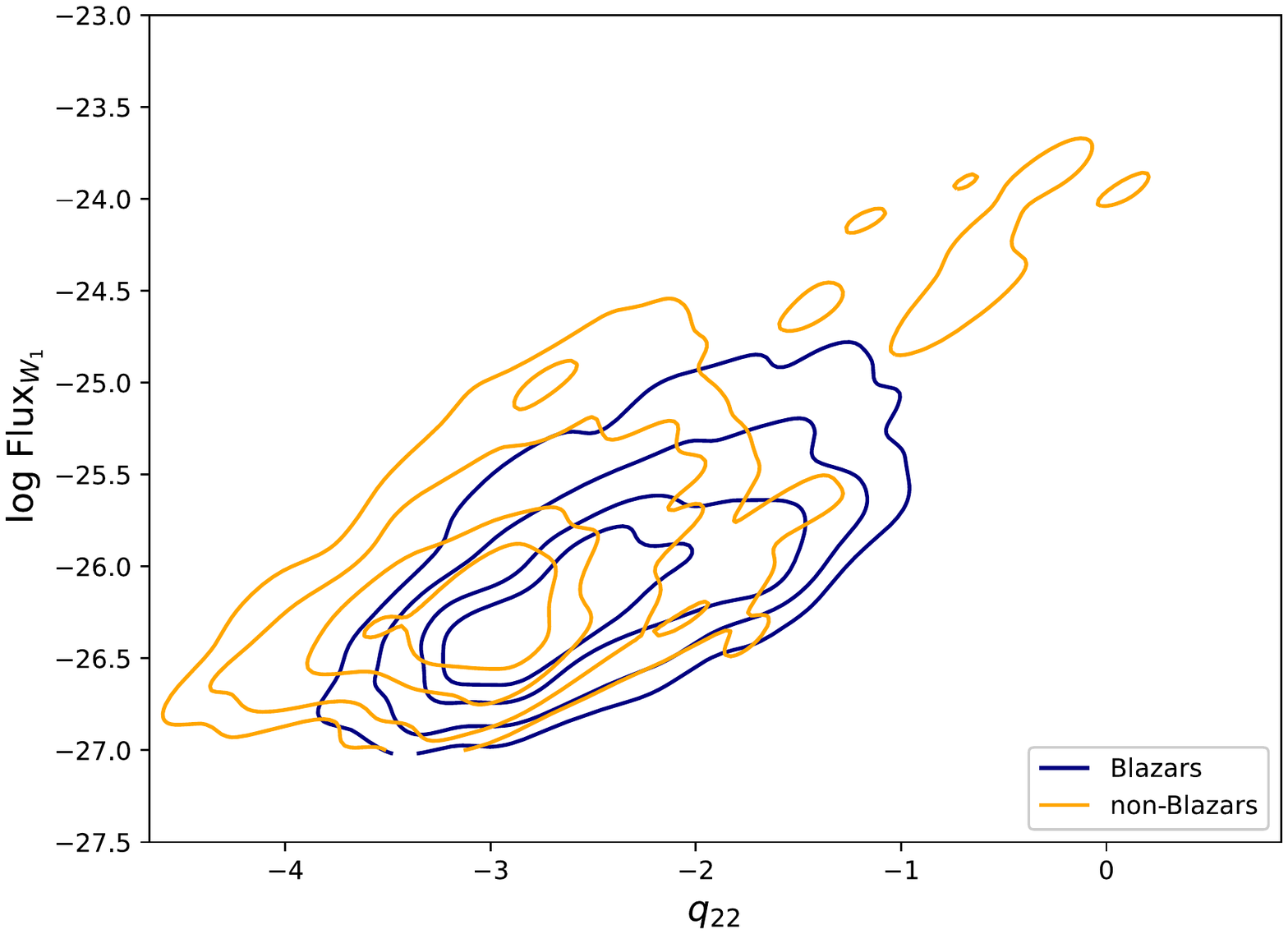}}
   \hfill
   \subfloat[]{\includegraphics[width=.33\textwidth,keepaspectratio]{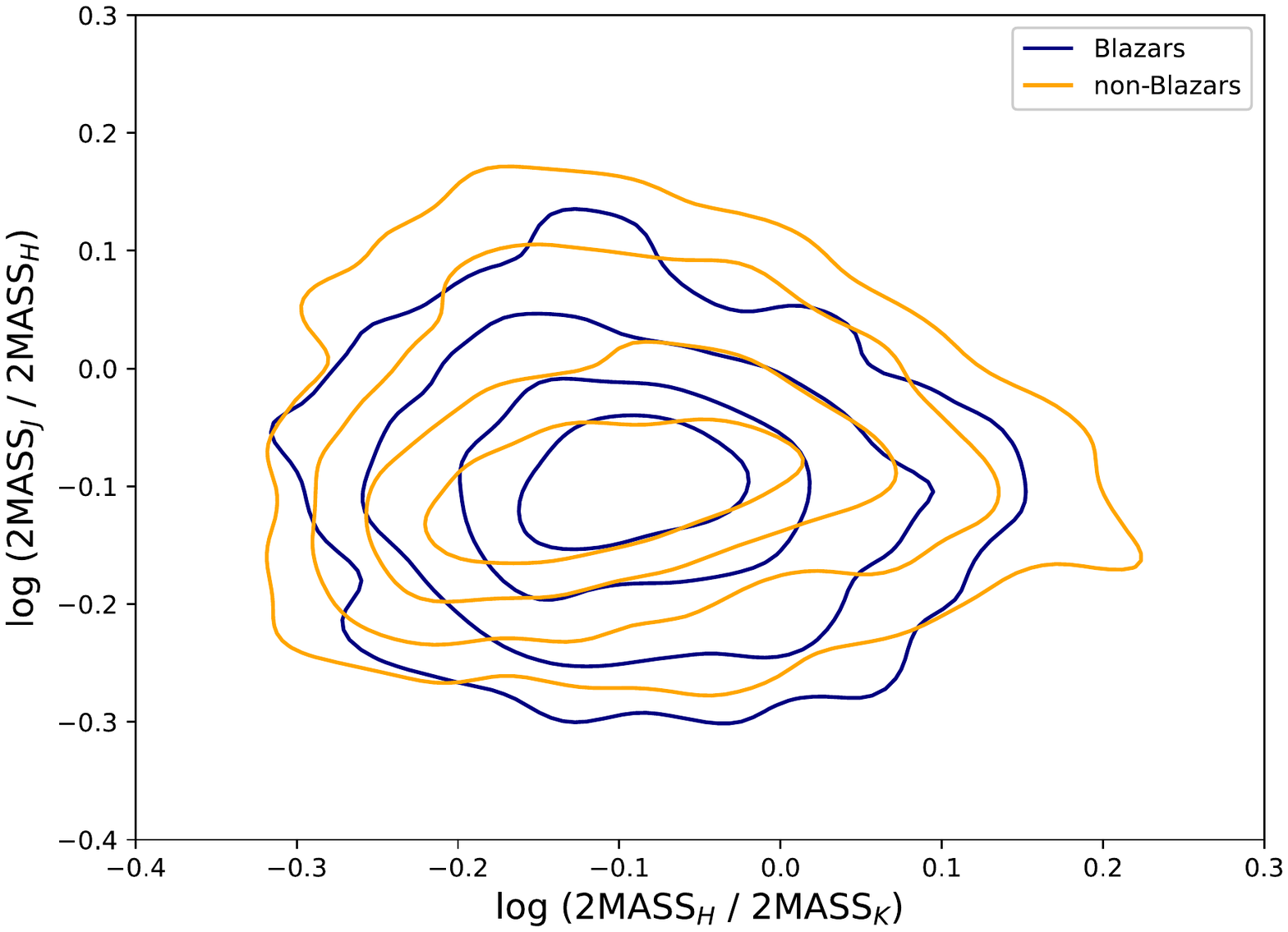}}
         
    \caption{The distributions of slopes and fluxes (smoothed by a gaussian kernel) for two combinations of parameters. (a) shows the slope between the two radio frequencies (1.4 GHz and 4.8 GHz) and the flux at 1.4 GHz, (b) shows $q_{22}$ parameter and the flux in the W1 band and (c) shows the difference in the logarithm of the flux in the 2MASS frequencies}. In both plots, the blue contours represent the Blazars, while the orange ones are for non-Blazars.
    \label{fig:parameters}
\end{figure*}

 Using the most important frequencies as determined by the net (listed in Table \ref{tab:important_frequencies}), we analysed if any combination could separate the classes.  Since the X-ray region is not important to the classification, a comparison with standard parameters to select Blazars (e.g. the $\alpha_{r-o}$ $\alpha_{o-x}$ plane for HBLs, the X-ray flux) is not possible. With the ten frequencies mentioned above, the total number of combinations of two or more parameters available is too large, so our search was not exhaustive; we focused mainly on the parameters mentioned above and a few others.  As expected, the slope in Radio provides some measure of separation due to radio-quiet sources in our dataset. However, no combination of the other parameters provided a separation at least as good as the radio slope. We present three combinations in Fig.~\ref{fig:parameters}, one with the Radio slope, one with a version of the q-parameter, using the flux in the W1 band at $3.4\mu$m and 1.4 GHz,  and the final one using 2MASS colors. 

\par Even though the analysis did not exhaust the whole parameter space, the fact that no combination of fluxes, slopes or colors was able to separate Blazars from non-Blazars validates our deep learning approach, where the Neural Network was able to utilize its more than nine million parameters to learn to distinguish between the two classes based solely on the SED.

% ====================================================================
% Section: Conclusion
% ====================================================================
\section{Discussion and concluding remarks}\label{sec:conclusion}
\subsection{Summary}
In this paper, we presented a deep learning model capable of classifying Blazars from other AGNs based only on their multiwavelength SED, achieving more than $97\%$ ROC AUC. 
\par To do this, we assembled a catalog of AGN SEDs using VOU-Blazars, a public tool developed in the context of the Open Universe Initiative and available through its online portal. We were able to obtain more than 14,000 multi-wavelength SEDs, potentially making it the most complete VO-based SED sample with publicly available data. This sample is extremely heterogeneous, with SEDs ranging from a few points to almost ten thousand, a challenge for deep learning models. We apply little preprocessing to the data, ensuring that the input is as close to real-world data as possible, making the model applicable to a variety of situations.
\subsection{Relevance and reliability of classification scheme presented}
\par Our deep model contains a bidirectional LSTM layer to take advantage of the fact that, although the SED peaks might move to lower or higher frequencies (or fluxes), the double-humped shape of a Blazar SED is maintained. 
\par A 10-fold cross-validation was done and the model performed similarly in all folds, with the reduced and the complete sample. We obtained mean AUCs for the ROC and Precision-Recall curve in a blind test sample above $0.97$ in both cases, with minimal variance between the folds. The confusion matrices show the net was able to correctly classify more than $90\%$ of the Blazars in both samples. These results show the net's robustness and its ability to generalize its training even with very heterogeneous datasets.
\par
\begin{figure*}
   \centering 
   \subfloat[ROC curve.]{\includegraphics[width=.49\textwidth,keepaspectratio]{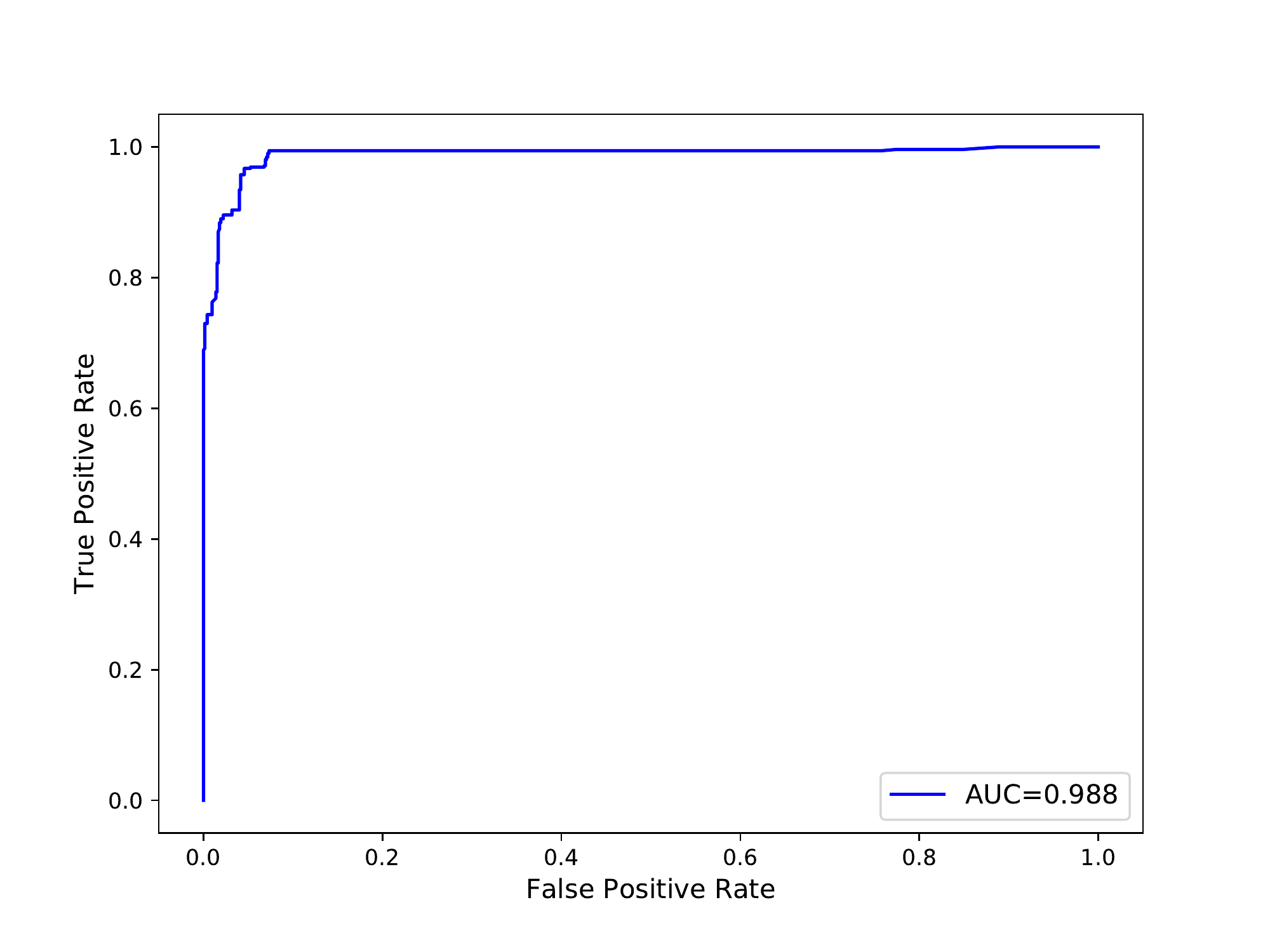}}
    \hfill
   \subfloat[Precision-Recall curve.]{\includegraphics[width=.49\textwidth,keepaspectratio]{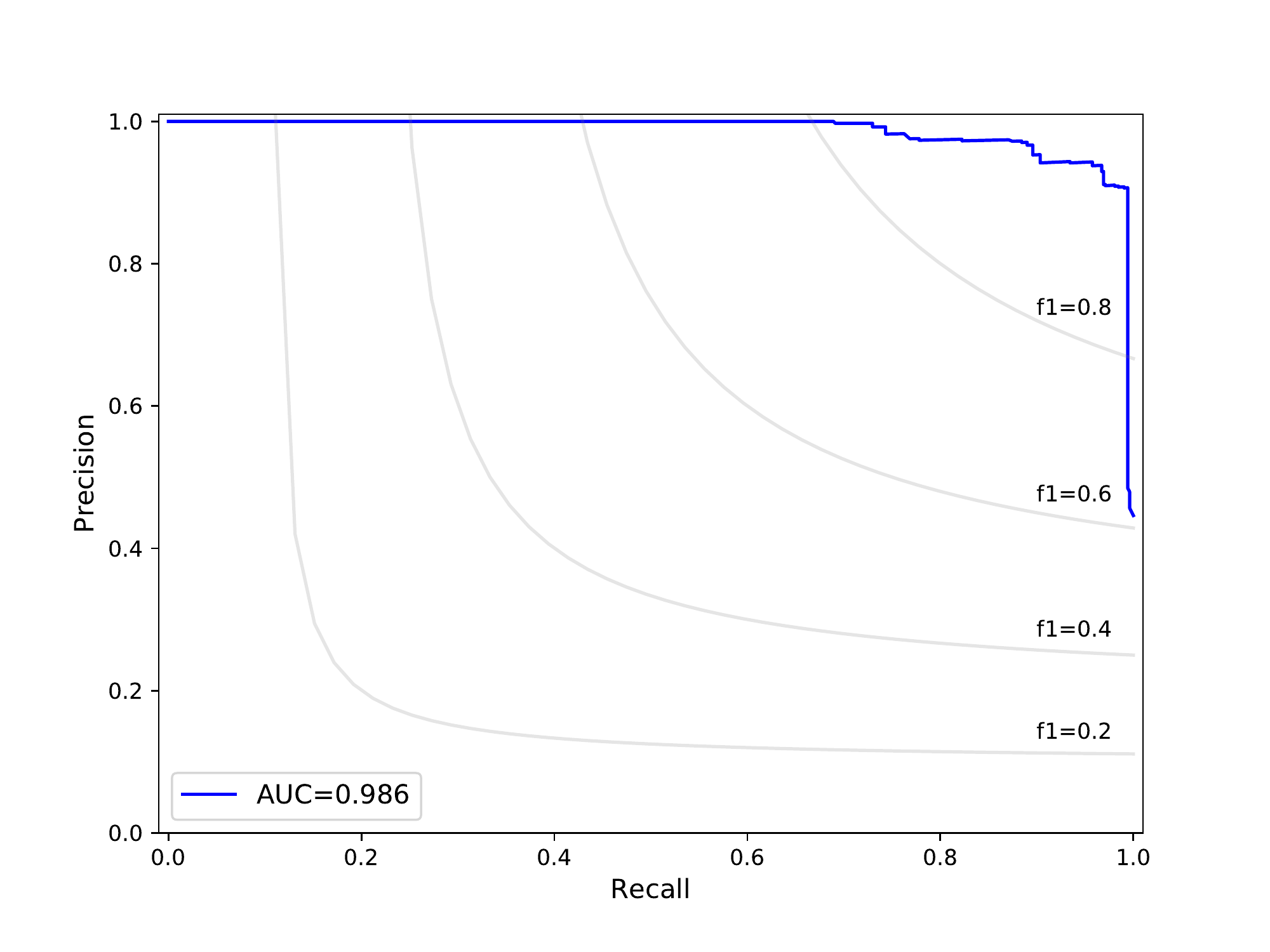}}
         
    \caption{ROC and Precision-Recall curves for the test set composed of the $10\%$ sparsest SEDs in our dataset. The model was trained using the other $90\%$, with $10\%$ of those used for validation. The results are compatible with the previous ones presented in Section~\ref{sec:results}}
    \label{fig:sparse_seds}
\end{figure*}

 We tested even more our model's robustness by using the $10\%$ sparsest SEDs as a testing set, keeping approximately the same distribution of blazars/non-blazars as in the full dataset. Even in this scenario, both the ROC and Precision-Recall AUC are compatible with our results within one sigma (see Figure~\ref{fig:sparse_seds}).

\par Despite its heterogeneity with regards to the SEDs, our dataset is comprised only of AGNs. However, it is still likely that the model will perform satisfactorily even with the inclusion of non-AGN sources since their SEDs, for the vast majority, are sufficiently different from that of blazars.
\subsection{Classification interpretation}
%* does the classification make sense to our intuition?
\par A connection between the net's results and the physics of AGNs can be attempted by analysing the gradients of a trained model, mapping the most important features to the classification to frequencies in the SED. With this, we identified the spectral regions most important to the classification of a source in blazar or non-blazar, recovering known results from observations. A search for parameters that could separate the classes could not find any combination, using the ten most important frequencies for both blazars and non-blazars, indicating that the separation is multi-parametric and validating our choice of a deep learning model. 
\subsection{Deep Learning perspective}

The high accuracy of the present model was derived using this open available data. However, some other samples might have a different coverage and could be highly incomplete in several frequencies. In such situations, the performance of the DL model could vary. We are currently investigating the use of models that make inferences with arbitrary number of frequencies and are prone to work with few data for each object.
%%%%%%%%%%%%%%%%%%%
\subsection{Application to Open Universe}
%%%%%%%%%%%%%%%%%%%%%
Given the relevance of blazars in contemporary high-energy multi-messenger astrophysics, it is important to assemble catalogues that are as large and complete as possible. 
The tool developed in this work can significantly contribute to this goal by finding new blazars in large samples of X-ray sources that also show radio emission. The ROSAT sky survey \citep[][]{RASS} covers the entire sky, but with a shallow X-ray sensitivity and probably all bright blazars included in this sample have already been identified. The ideal deep catalogue for this purpose would be that generated by the SRG/eROSITA all-sky X-ray survey \citep{e-rositabook,e-rosita2020}, which is currently on-going and will be publicly available in the medium term. Meanwhile, existing X-ray samples such as OUSXG \citep{OUXG}, 2SXPS \citep{2sxps}, 4XMM-DR9 \citep{4xmm}, CSC2 \citep{csc2}, and XMMSL2 \citep{XMMSLEW}, which collectively cover about over 20\% of the sky with sensitivities ranging from below 10$^{-15}$ to a few times 10$^{-13}$ \ergs, are deep and wide enough to detect a large number of new blazars. These could be preliminarily identified (and used as input to our tool) by cross-matching the X-ray sources with large area catalogs of radio sources, such as the NVSS \citep{NVSS}, with a sensitivity of $\sim$ 2.5 mJy @ 1.4GHz, VLASSQL ($\sim$ 1 mJy @ 3GHz) \citep{VLASSQL} and SUMSS21 ($\sim$ 5 mJy @ 0.8GHz) \citep{SUMSS}.
A rough estimate of the number of blazars detectable with this method can be obtained from the radio LogN-LogS of blazars of Chang et al. (2021, to be submitted), considering the radio flux densities that, for different types of Blazars, correspond to X-ray fluxes at the typical sensitivity limit of the available X-ray catalogs. Considering a $\nu F_{\nu}$ flux at 1 keV of 5$\times 10^{-14}$ \ergs \citep[see Fig. 7 of ][]{Giommi2020} as a typical X-ray sensitivity of the deeper existing catalogues, this corresponds to a radio flux density of $\sim $30 mJy for FSRQs and LBL BL Lacs, $\sim$ 15 mJy, for IBL sources and $<$ 1 mJy for HBL objects. The corresponding space densities are 3, 0.5, and 0.3 sources per square degree, respectively. For an area of sky covered of 3,000 square degrees (at the assumed sensitivity), this gives 9,000 LBLs, 1,500 IBLs, and 900 HBLs. The number of HBLs is actually an underestimation since the X-ray flux in these sources can be two or even three orders of magnitudes higher than that of LBLs and IBLs for the same radio intensity, and therefore, these sources can also be detected where the X-ray sensitivity is of the order of $10^{-12}$ \ergs\, which is much larger than 3,000 square degrees considered above.
However, these numbers should be taken with caution, as large variability that characterises blazars, and the order-of-magnitude approximations used make these numbers uncertain. Even in a conservative scenario, the number of new blazars that can be discovered with this method is much larger than the combined number of known blazars in the 5BZCAT, 3HSP and 4LAC catalogs.

% perspectives

Another interesting application of our tool would be the identification of candidate transient blazars, objects that normally are not detectable in the X-ray or gamma-ray bands but are flat spectrum radio sources with infrared data identical to that of HBL blazars and occasionally flare to become bright HBL sources with large X-ray fluxes and detectable gamma-ray emission. At the moment, only the case of 4FGLJ1544.3-0649 is known \citep{TransientBlazar}, a blazar which for a few months rose to be one of the brightest known X-ray blazars, but it is possible that these objects are relatively common and play a non-negligible role in blazar demographics and possibly in multi-messenger astrophysics. 

%=========================================
\section*{Data Availability}
All SEDs were obtained using the VOU-Blazars tool, available at the Open Universe online portal. All data recovered with it is also publicly available either via VO or Open Universe. 
% ====================================================================
% Section: Acknowledgements
% ====================================================================
\section*{Acknowledgements}
We would like to thank Luciana Dias for the help with the early stages of tis work.
\par
\textbf{UBdA} acknowledges a Serrapilheira Institute Grant no. Serra - 1812.26906 and a FAPERJ Young Scientist Fellowship no. E-26/202.818/2019. He also acknowledges a CNPq Research Productivity Grant no. 311997/2019-8. \\
\textbf{PG} acknowledges the support of the Technische Universit\"at M\"unchen - Institute for Advanced Studies, funded by the German Excellence Initiative (and the European Union Seventh Framework Programme under grant agreement no. 291763) and the support by the Excellence Cluster ORIGINS, which is funded by the Deutsche Forschungsgemeinschaft (DFG, German Research Foundation) under Germany's Excellence Strategy -EXEC-2094 - 390783311.\\
The authors made use of multi GPU Sci-Mind machines developed and tested for Artificial Intelligence and would like to thank Marcelo P. Albuquerque, P. Russano and P. Souza Pereira for the infrastructure support. This paper also made use of the Plot Deep Design \footnote{\url{https://github.com/cdebom/plot_deep_design}} library to make plots of the presented architecture. 
We acknowledge the use of data, analysis tools and services from the Open Universe platform, the Astrophysics Data System (ADS), and the National Extra-galactic Database (NED).
% ====================================================================
% Section: Bibliography
% ====================================================================
\newpage
\bibliographystyle{mnras} 
\typeout{}
\bibliography{bibliografia}

\bsp
\label{lastpage}

\end{document}